# Cavity-mediated thermal control of metal-to-insulator transition in 1T-TaS$_2$

Giacomo Jarc[1,2], Shahla Yasmin Mathengattil[1,2], Angela Montanaro[1,2,3], Francesca Giusti[1,2], Enrico Maria Rigoni[1,2], Rudi Sergo[2], Francesca Fassioli[3,4], Stephan Winnerl[5], Simone Dal Zilio[6], Dragan Mihailovic[7], Peter Prelovšek[7], Martin Eckstein[8], and Daniele Fausti[1,2,3].

[1]*Department of Physics, Università degli Studi di Trieste, 34127 Trieste, Italy*
[2]*Elettra Sincrotrone Trieste S.C.p.A., 34127 Basovizza Trieste, Italy*
[3]*Department of Physics, University of Erlangen-Nürnberg, 91058 Erlangen, Germany*
[4]*International School for Advanced Studies (SISSA), Via Bonomea 265, I-34136 Trieste, Italy*
[5]*Institute of Ion Beam Physics and Materials Research, Helmholtz-Zentrum Dresden-Rossendorf, Bautzner Landstrasse 400, 01328, Dresden, Germany*
[6]*CNR-IOM TASC Laboratory, Trieste 34139, Italy*
[7]*Jožef Stefan Institute, Jamova 39, 1000 Ljubljana, Slovenia*
[8]*Institute of Theoretical Physics, University of Hamburg, Notkestrasse 9, 22607 Hamburg, Germany*

*Correspondence: daniele.fausti@elettra.eu*

**Placing quantum materials into optical cavities provides a unique platform for controlling quantum cooperative properties of matter, via both weak and strong light-matter coupling [1,2]. Here we report the experimental evidence of reversible cavity control of a metal-to-insulator phase transition in a correlated solid-state material. We embed the charge density wave material 1T-TaS$_2$ into cryogenic tunable terahertz cavities [3] and show that a switch between conductive and insulating behaviors, associated with a large change in the sample temperature, is obtained by mechanically tuning the distance between the cavity mirrors and their alignment. The large thermal modification observed is indicative of a Purcell-like scenario in which the spectral profile of the cavity modifies the energy exchange between the material and the external electromagnetic field. Our findings provide opportunities for controlling the thermodynamics and macroscopic transport properties of quantum materials by engineering their electromagnetic environment.**

## MAIN

Optical driving with ultrashort pulses has been extensively used to dynamically control the properties of complex quantum materials [4-9]. Yet, several theoretical proposals indicate that the control of materials functionalities can be obtained by modifying their electromagnetic environment, embedding the materials into optical cavities, even in absence of a driving field [1,2]. Predictions range from enhanced superconductivity through cavity-mediated electron pairing [10-15], cavity control of the competing order between charge density wave and superconducting phases [16], cavity control of excitons [17], enhanced ferroelectricity [18-20], and cavity control of magnetic orders [21]. Experimentally, it has been demonstrated that vacuum fields in the strong coupling regime [1] can change material functionalities as, for example, the magneto-transport in two-dimensional materials [22], the topological protection of the integer quantum Hall effect [23], or the ferromagnetic order in unconventional superconductors [24].

Cavity control of phase transformations in complex systems can be achieved via distinct physical mechanisms. On the one hand, the selective coupling of the cavity modes to the excitations of a given phase can renormalize its free energy with respect to that of other ones, thereby modifying the temperature at which the phase transition occurs. On the other hand, a cavity can reshape the exchange of energy between a material and the thermal reservoir of photons in which the material is immersed [25]. By engineering the density of states of the electromagnetic environment at the sample position through tunable optical cavities, it is therefore possible to modify the sample's absorption and emission [26-29], and, in turn its temperature ($T_{int}$). Fig. 1A illustrates the two aforementioned cavity-mediated mechanisms.

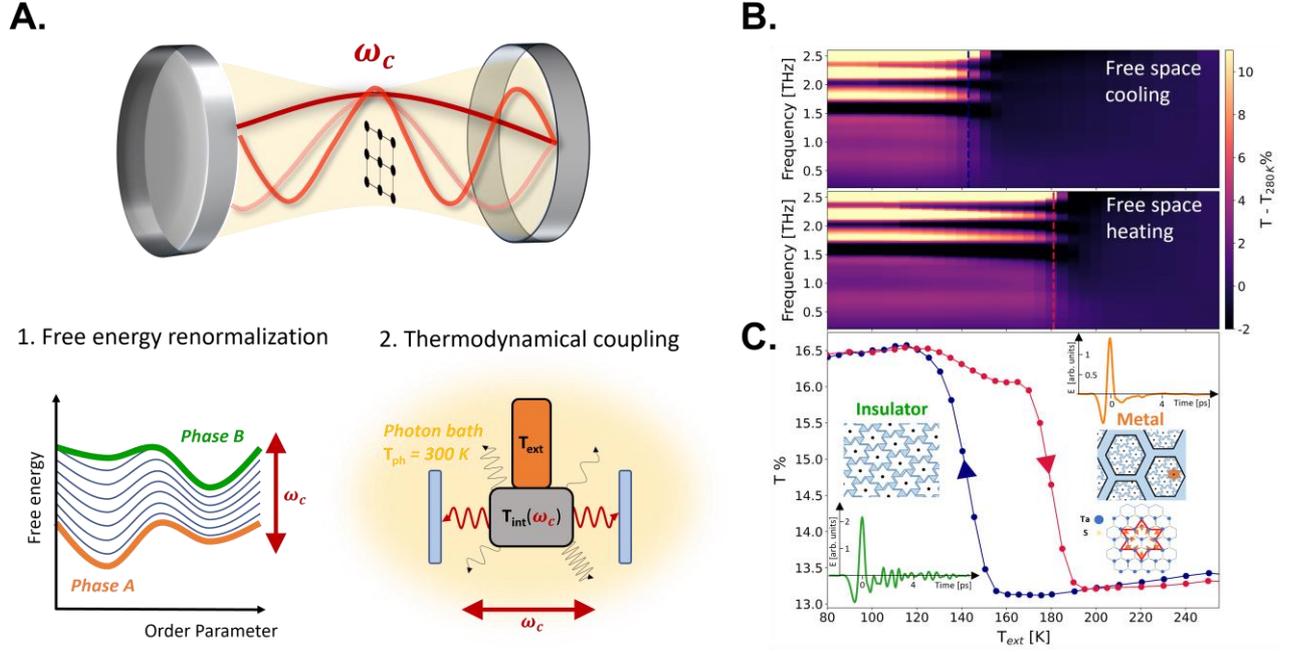

**Fig. 1**: **Mechanisms of cavity control of quantum material states and THz characterization of 1T-TaS$_2$ metal-to-insulator transition**. **A.** Schematic of a material embedded in the middle of a tunable optical cavity with controllable fundamental frequency $\omega_c$ and alignment. Coupling of the material's excitations with the cavity mode can act on the thermodynamics of the sample within two different scenarios. On the one hand, it can renormalize the free energy of one material phase with respect to the other (bottom left panel). On the other hand, as a function of $\omega_c$ the cavity can reshape the emission and absorption of the material, subsequently rescaling its local temperature $T_{int}(\omega_c)$ with respect to the temperature measured on the sample support ($T_{ext}$). **B.** THz linear transmission spectra in free space at different temperatures across 1T-TaS$_2$ metal-to-insulator transition (temperature scans performed by cooling (upper panel) and heating (lower panel)). To highlight the phase transition, each spectrum has been subtracted from the 280 K THz transmission. **C.** Temperature dependence of the integrated low frequency transmission (0.2 THz < $\omega$ < 1.5 THz), marking the metal-to-insulator transition and its hysteresis. In the insets the time domain THz fields are shown for the metallic and the insulating phases, together with the illustration of the in-plane lattice modulations characteristic of the insulating C-CDW phase and of the metallic NC-CDW phase.

In this work, we investigate the metal-to-insulator phase transition in the transition metal dichalcogenide 1T-TaS$_2$ embedded in low-energy terahertz (THz) and sub-THz cryogenic cavities (Fig. 1A). 1T-TaS$_2$ exhibits a temperature-dependent charge order that originates from the competition of Coulomb repulsion, lattice strain, interlayer hopping, and Fermi surface nesting [30-32]. At ambient temperature, 1T-TaS$_2$ is in a nearly commensurate charge density wave (NC-CDW) phase with metallic character, featuring hexagonal-shaped polaron domains [30,33,34] forming a David's star pattern [35-37] (Fig. 1C). By lowering the temperature below ~ 180 K, a transition to an insulating commensurate charge density wave (C-CDW) state occurs [31,38]. We note that the free energy landscape of 1T-TaS$_2$ is much more complicated than the simple sketch in Fig. 1A: the phase transitions in 1T-TaS$_2$ are multiple and sensitive to the thermal history of the sample. Upon heating from the C-CDW phase, an additional intermediate trigonal (T) phase with in-plane charge stripes occurs at around 220 K and persists up to around 280 K, when the NC-CDW is re-established [39].

THz spectroscopy is a powerful tool for tracking the metal-to-insulator transition since it is able to measure contactless the quasi-static dielectric response associated to the presence of conductive charges characteristic of a metallic state (see Methods and Ref. [3] for further details on the experimental set-up). Here we employ broadband time-domain THz spectroscopy to track the charge order in the sample for different cavity settings. We demonstrate that a bidirectional switch between the metallic and insulating phase can be obtained by tuning the cavity length and by adjusting the alignment of its mirrors while keeping the cryogenic temperature of the sample support and mirrors fixed.

It is important to highlight that a simultaneous measurement of the actual sample's temperature inside the cavity ($T_{int}$) and THz transmission is not viable. At a practical level, the placement of a physical thermometer within the cavity would absorb the THz pulses and make the transmission measurements unfeasible. At a fundamental level, any object placed within the optical cavity will perturb the cavity environment and therefore the response of the light-matter assembly. For this reason, we have designed an experimental protocol in which for the THz characterization we measure the temperature on the cold-finger support of the sample outside of the cavity, denoted by $T_{ext}$ (Fig. 1A). This protocol allows us to identify an effective critical temperature $T_c^{eff}$ for the phase transition which is defined as the temperature of the support at which the phase transition is observed. In a separate measurement campaign, we place a micrometric thermocouple, and we measure for different experimental configurations the temperature at the sample position (with and without the sample) which we denote as $T_{int}$, while simultaneously monitoring the external temperature $T_{ext}$.

**THZ SPECTROSCOPY OF 1T-TaS$_2$**

Figure 1B shows the THz linear transmission of 1T-TaS$_2$ in free space upon heating and cooling as a function of the temperature of the sample support ($T_{ext}$). This captures the first-order transition between the NC-CDW metallic phase and the C-CDW insulating phase. The phase transition results in: i) an increase of the low frequency transmission (0.2 THz < ω < 1.5 THz) below the effective critical temperature, which is consistent with a transition to an insulating behavior (Drude-like response of free carriers vanishes in the insulating phase [40,41]); and ii) the emergence below $T_c^{eff}$ of infrared-active optical phonons at 1.58 THz, 2.04 THz and 2.35 THz, which are screened by the free carriers and therefore not visible in the metallic phase (Fig. 1C, insets, report time-domain THz traces representative for the two phases). We will use the temperature dependence of the integrated low-frequency transmission (0.2 THz < ω < 1.5 THz) as a marker which tracks the charge order dynamics in 1T-TaS$_2$ and hence the metal-to-insulator phase transition (Fig. 1C). The low frequency transmission is directly mapped into the evolution of the Drude optical conductivity $\sigma_1(\omega)$ representative of the free carriers response (Supplementary Information). Analogous transition temperatures can be obtained by tracking the temperature dependence of the transmission at the phonon frequency (Methods). The temperature dependence of the integrated low-frequency transmission (0.2 THz < ω < 1.5 THz) for the material in free space is shown in Fig. 1C. The difference between the results obtained upon heating and cooling the sample in free space marks the hysteresis associated to the first-order phase transition. The phase transition in free space upon heating occurs at $T_{ext}$ = 181 K and at 143 K upon cooling from the metallic phase. Note that the smooth transition observed can be ascribed to the presence of intrinsic inhomogeneities and strain in the system, which may smear out the first-order transition [42-44] (Methods). The effective critical temperature $T_c^{eff}$ measured in our set-up differs from the literature value [31] by about 35 K. This discrepancy is attributed to the difference between the internal temperature of the sample ($T_{int}$) and the temperature of the cryostat's cold finger ($T_{ext}$), as a consequence of the small thermal conductivity of the silicon nitride membranes holding the 1T-TaS$_2$ sample [3]. A finite-elements simulation of the membrane's thermal profile is in quantitative agreement with the measured temperature shift (details in the Methods section).

## CHARACTERIZATION OF 1T-TaS$_2$ IN CRYOGENIC THz FABRY PÉROT CAVITIES

Figure 2 presents the THz linear transmission as a function of the sample holder temperature (cooling in Fig. 2A, and heating in Fig. 2B) of 1T-TaS$_2$ in free space and embedded in the center of an optical cavity with resonant frequency $\omega_c$ = 11.5 GHz and quality factor $Q \sim 4$ (Methods). The placement of the sample in such a cavity results in a modification of $T_c^{eff}$ for the metal-to-insulator transition, which is observed at 136 K on heating and at 109 K on cooling. The modification of $T_c^{eff}$ depends also on the thermal cycle. Indeed, a change of $T_c^{eff}$ of 44 K is observed if the critical temperature is approached from the insulating state (heating), while a shift of 33 K is obtained starting from the metallic phase (cooling), resulting in a shrinking of the hysteresis of about 11 K. We highlight that $T_c^{eff}$ is independent from the input intensity of the THz field, which, therefore, acts only as a probe and does not introduce a detectable thermal load on the sample (see Supplementary Fig. S15 for measurements with different THz field strengths).

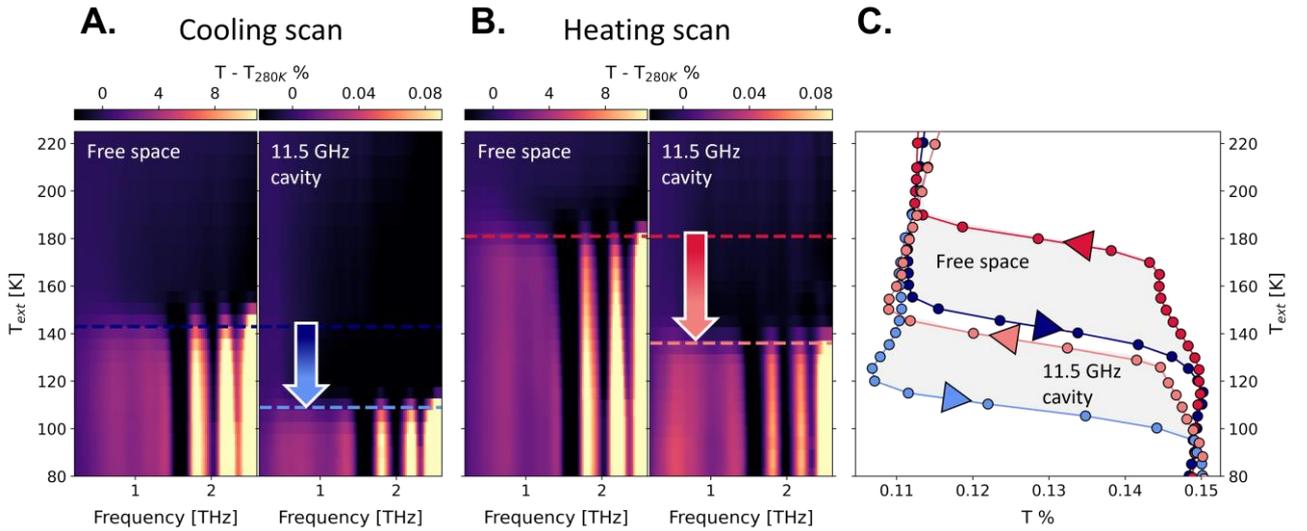

**Fig. 2**: **Renormalization of the effective critical temperature of the metal-to-insulator phase transition within the cavity. A**, **B.** Temperature-dependent THz transmission upon cooling (**A**) and heating (**B**) for a sample held in free space (left) and one placed in the middle of the 11.5 GHz cavity (right). **C.** Comparison between the hysteresis in free space and within the 11.5 GHz cavity plotted as the integrated cavity transmission in the range 0.2 THz < ω < 1.5 THz. The free-space data have been arbitrarily translated along the horizontal axis to overlap with the cavity integrated transmission. In the cavity, a renormalization of the effective critical temperature of 44 K (33 K) towards lower temperatures is measured upon heating (cooling) the sample. This results in a shrinking of the effective phase-transition hysteresis of 11 K within the cavity.

Next, we varied the cavity geometry and measure $T_c^{eff}$ as a function of the alignment of the cavity mirrors. We quantify the cavity misalignment as the sum of the misalignment angles of the two cavity mirrors Θ with respect to the parallel mirrors configuration. The temperature dependence of the low-frequency THz transmission (0.2 THz < ω < 1.5 THz integration range) at different mirror alignments is shown in Fig 3A for the temperature scans performed by heating and cooling the cold-finger sample holder. Misaligning the mirrors modifies $T_c^{eff}$, which approaches the free space value when the cavity is highly misaligned (Fig. 3A). In the inset of Fig. 3A, we show that a switch between the metallic and the dielectric linear response is obtained at fixed $T_{ext}$ by solely changing the cavity alignment. As any misalignment of the cavity mirrors reduces the photon lifetime within the cavity (and hence the quality factor), the sensitivity of $T_c^{eff}$ not only to the presence of the cavity but also to the mirrors alignment, is suggestive of a cavity-mediated effect. This is further supported by the fact that misaligning the cavity mirrors not only changes the effective critical temperature, but also increases the effective hysteresis of the metal-to-insulator transition towards its free space value.

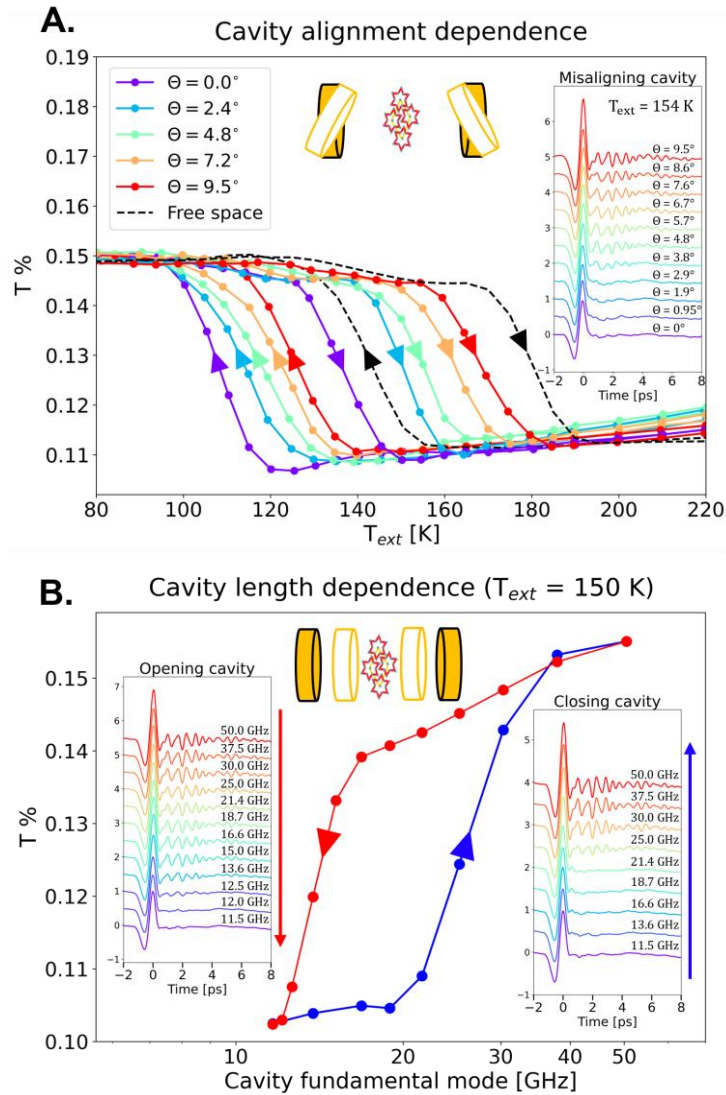

**Fig. 3**: **Dependence of the effective critical temperature on the cavity geometry**. **A.** Dependence of the effective metal-to-insulator phase transition as a function of the cavity alignment for the 11.5 GHz cavity. The hysteresis is plotted for each misalignment angle Θ as the integrated low frequency transmission (0.2 THz < ω < 1.5 THz). In the inset panel, the THz fields detected at the output of the coupled 11.5 GHz cavity at fixed temperature ($T_{ext}$ = 154 K) as a function of the mirrors alignment. Transition from the dielectric to the metallic behavior is detected passing from the misaligned to the aligned configuration. **B.** Reversible cavity control of the metal-to-insulator transition at fixed temperature ($T_{ext}$ = 150 K). The hysteresis as a function of the cavity fundamental mode is plotted as the evolution of the integrated low frequency THz transmission (0.2 THz < ω < 1.5 THz). The insets show the evolution of the time domain THz fields transmitted for different values of the cavity frequency ranging from 50.0 GHz to 11.5 GHz (opening cavity case) and from 11.5 GHz to 50.0 GHz (closing cavity case), demonstrating the reversible switching between the two phases.

Figure 3B reports the THz transmission as a function of the cavity fundamental frequency at a fixed cold-finger temperature ($T_{ext}$ = 150 K). The results reveal that the cavity-mediated change of $T_c^{eff}$ overcomes the free space hysteresis, thus enabling a reversible touchless control of the metal-to-insulator phase transition. Upon reducing the distance between the mirrors, we detected the phase transition between the metallic and the insulating phase to occur at a cavity frequency of 25.0 GHz. This is highlighted in the THz time domain traces of the insets of Fig. 3B by the screening of the infrared-active phonon modes of the C-CDW insulating phase. After the system has fully switched to the insulating state, we decreased the cavity fundamental frequency and detected a switch to the metallic phase at a lower cavity frequency (13.6 GHz). This results in the cavity frequency-dependent hysteresis highlighted in Fig. 3B.

The detailed dependence of $T_c^{eff}$ on the cavity fundamental mode is presented in Fig. 4A. We measured the effective transition temperature for cavity frequencies ranging from 11.5 GHz until 570 GHz. Importantly, the maximum cavity frequency employed lies below the frequency of the lowest IR-active mode of the C-CDW phase (1.58 THz as shown in Fig. 1C). We made this choice in order to disentangle possible effects due to the coupling to the infrared-active optical phonons of the CDW. Figure 4A summarizes the dependence of the measured $T_c^{eff}$ (heating and cooling) on the cavity resonant frequency. The results show a non-monotonic trend of $T_c^{eff}$ as a function of cavity resonant frequency with respect to the free-space condition. Indeed, while long wavelength cavities (up to ~ 25 GHz) stabilize the nearly-commensurate metallic phase, higher energy cavities effectively favor the insulating C-CDW phase with respect to the material in free space (represented in the figure as the zero-frequency point). Overall, we revealed a shift of $T_c^{eff}$ of 75 K by moving from the lowest energy cavity employed in the experiment (11.5 GHz) towards the highest one (570 GHz).

We note that the value of $T_c^{eff}$ in the cavity with the lowest frequency achievable in our set-up is approximatively 30 K below the $T_c^{eff}$ measured in free space (Fig. 4A). This anomalous behavior cannot be rationalized by incoherent radiation heating due to the presence of the mirrors. Crucially, the exclusion of an incoherent heating mechanism is confirmed by the fact that the dependence of $T_c^{eff}$ on the cavity geometry is qualitatively similar for measurements with cavity mirrors at different temperatures (Supplementary Figs. S4 and S5). This is in stark contrast with a scenario in which thermal radiation is transmitted incoherently to the sample (incoherent radiation heating), which would give opposite trends with hot and cold cavity mirrors (see finite-elements simulation of incoherent radiation heating with hot and cold mirrors in Supplementary Figs. S6 and S7 and the discussion therein).

Having established that the observed effect cannot be rationalized by incoherent radiation heating, in the following we focus on understanding if the observation could be explained by a cavity-mediated heating (cooling) or by a free energy renormalization (the two scenarios presented in the introduction, Fig. 1A). To determine whether the cavity is influencing the temperature at the sample position (i.e., inside the cavity), we performed an independent measurement campaign to simultaneously measure $T_{int}$ and $T_{ext}$. To this purpose, a micrometric Cr-Al junction custom-designed was used (Methods).

Figure 4B shows, for representative cavity frequencies, the difference between the temperature measured within the cavity ($T_{int}$, in thermal contact with the sample) and the temperature of the cold finger ($T_{ext}$) as a function of $T_{int}$. Upon changing the cavity mode, we revealed a non-monotonic trend of the differential temperature $T_{int} - T_{ext}$ with respect to the free-space configuration. Whereas lower frequency cavities induce a coherent heating of the sample, the coupling with higher energy cavity modes decreases the temperature of the sample with respect to the free space conditions. By tracking the $T_{ext}$ at which $T_{int}$ = 210 K (nominal critical temperature of 1T-TaS$_2$) we revealed a non-monotonic trend as a function of the cavity frequency (Fig. 4B, inset), which is qualitatively consistent with the observation in Fig. 4A.

It is important to stress two aspects. i) The observed anomalous non-monotonic trend of $T_{int} - T_{ext}$, as well as its dependence on the cavity frequency, depends on the presence of the sample (the temperature difference is much smaller and monotonic when the thermocouple is mounted within the membranes without the sample (Supplementary Fig. S1A, B); ii) We repeated the measurements with mirrors at 290 K, revealing, apart from a rigid shift, a trend with cavity frequency analogous to the one observed with cryogenic mirrors (Supplementary Fig. S8). Crucially, a decrease in $T_{int}$ on closing the cavity, regardless of the mirrors temperature, is incompatible with an incoherent radiative scenario and points towards a cavity-mediated effect.

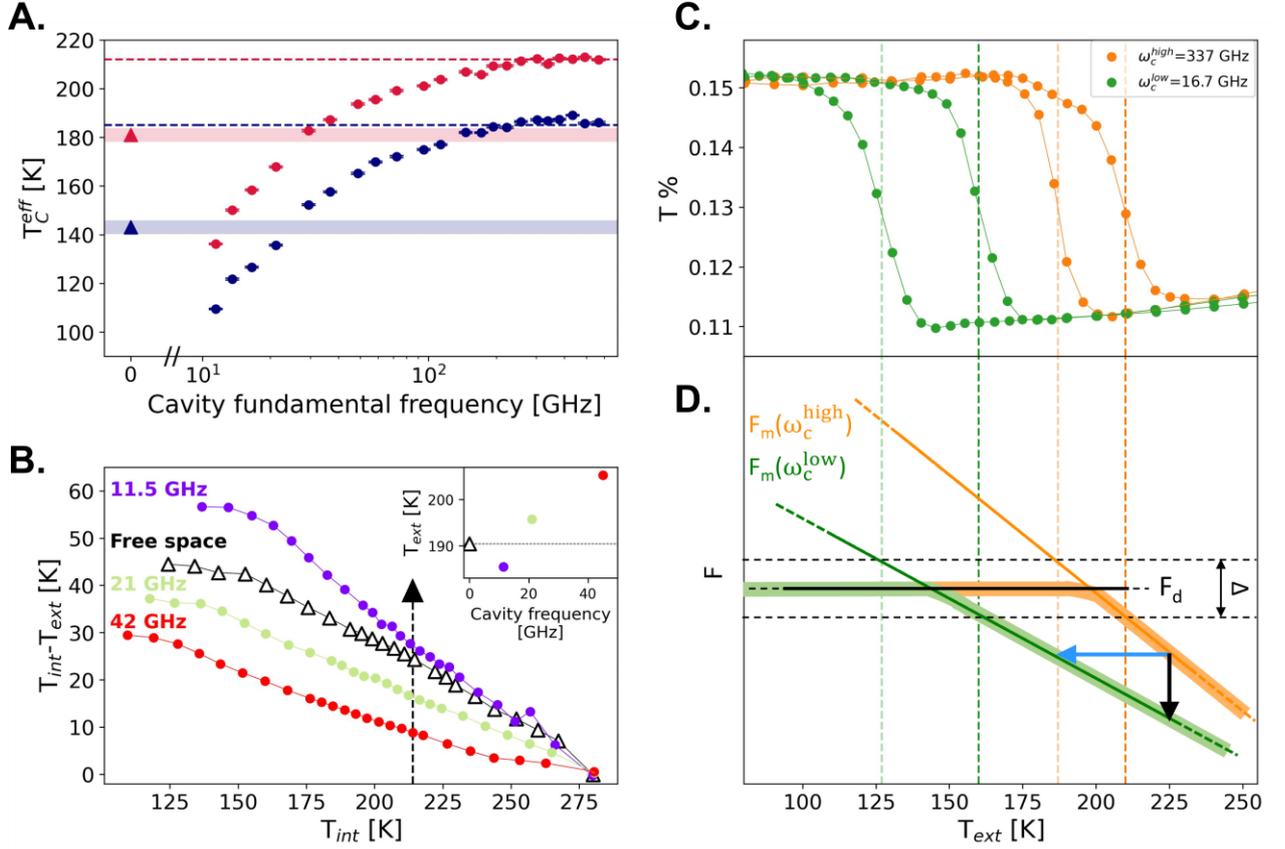

**Fig. 4**: **Cavity-mediated thermodynamics of the metal-to-insulator phase transition in 1T-TaS$_2$. A.** Dependence of the effective critical temperature on the cavity fundamental frequency for the heating and cooling temperature scans. The zero-frequency point represents the effective critical temperature measured in free space; the red (blue) dashed line indicates the literature critical temperature [31] for the heating (cooling) temperature scan. The error bar associated to each temperature is the standard deviation of the effective critical temperatures estimated for three consecutive scans. **B.** Difference between the temperature measured on the sample ($T_{int}$) and on the cold finger ($T_{ext}$) as a function of the sample temperature for different values of the cavity fundamental frequency. Temperatures have been measured upon heating the sample from the C-CDW phase. In the inset, the effective critical temperature, defined as the cold-finger temperature at which the sample reaches the nominal critical temperature $T_{int}$ = 215 K, as a function of the cavity mode. **C.** Comparison of the phase transition hysteresis of 1T-TaS$_2$ within a low-frequency cavity ($\omega_c^{low}$ = 16.7 GHz) and a high-frequency cavity ($\omega_c^{high}$ = 337 GHz). **D.** Schematic temperature dependence of the free energy of the metallic ($F_m$) and the dielectric ($F_d$) phase at the cavity frequencies $\omega_c^{low}$ and $\omega_c^{high}$ employed in C. The activation energy for switching the phase is indicated with Δ. The shift of the apparent transition temperature could be rationalized with a cavity-mediated renormalization of the free energy of the metallic state (black vertical arrow) or with a scaling of the sample effective temperature in analogy with the Purcell effect (blue horizontal arrow).

In light of the temperature measurements of Fig. 4B, the cavity frequency-dependent hysteresis in Fig. 3B can be rediscussed in terms of an effective renormalization of the sample temperature in the presence of the cavity. By keeping fixed the cold-finger temperature ($T_{ext}$), the local temperature of the sample ($T_{int}$) decreases upon increasing the cavity frequency (see for example Fig. S8). Therefore, closing the cavity effectively corresponds to cooling down the sample that is thus driven to the insulating state (blue curve in Fig. 3B). The effect is reversed when, starting from the insulating state, the cavity frequency is decreased (red curve). Closing and opening the cavity, leading to the hysteretic behavior in Fig. 3B, can then be interpreted as an effective change of the sample temperature induced by the cavity environment which is different in the two phases. Similarly, the alignment dependence of Fig. 3A can be linked to a mirror-controlled change of the temperature of the cavity-confined material (Supplementary Fig. S9).

Changing the cavity frequency leads not only to a renormalization of $T_c^{eff}$, but also to an effective shrinking of the hysteresis of the phase transition. This is demonstrated in Fig. 4C where we plot the comparison of the phase transition hysteresis for a low- and high-frequency cavity ($\omega_c^{low}$ = 16.7 GHz, and $\omega_c^{high}$ = 337 GHz). The measured changes in the effective critical temperature upon heating and cooling depend therefore not only on the cavity length, but also on the sample's thermal history (see Supplementary Fig. S2 and related discussion in SM). The observed modification of the cavity hysteresis hints at a possible scenario in which the coupling between 1T-TaS$_2$ and the cavity modes is the driving force of the effective renormalization of the transition temperature. Indeed, such coupling is expected to be different in the two phases, displaying a profoundly different dielectric response.

## DISCUSSION

The cavity-dependent effect on $T_c^{eff}$ can be rationalized within a thermodynamical picture considering how the free energies of the metallic phase ($F_m$) and of the dielectric phase ($F_d$) vary with the external temperature for different cavities (Fig. 4D). The crossing temperature between the free energies of the two phases sets the critical temperature of the metal-to-insulator transition. For simplicity, we assume $F_d$ to be weakly dependent on temperature and cavity geometry, and subsequently consider the temperature dependence of $F_m$ to be responsible for the phase transition. Figure 4D shows the schematic temperature dependence of $F_m$ and $F_d$ in low- and high-frequency cavities, consistent with the experimental observation in Fig. 4C.

This thermodynamical picture can be connected to the two proposed mechanisms leading to a cavity-induced modification of $T_c^{eff}$ discussed in Fig. 1A. In the first scenario, the coupling of the cavity modes with the sample changes the energy spectrum of the collective modes in the sample. In this picture, the cavity induces a decrease of the free energy of the metallic phase with respect to the insulating one, resulting in an effective shift of the observed transition temperature (black vertical arrow in Fig. 4D). The experimental observation would thus suggest that lowering the cavity resonance could cause a decrease in the free energy of the metallic phase and a reduction in the slope of its temperature dependence, consistent with a cavity-driven shrinking of the hysteresis.

To test whether the free energy scenario could be consistent with the experimental trends, we resort to a Dicke-based model with a single cavity mode coupled to a continuum absorption spectrum within the GHz spectral range, where conductivity measurements suggest an increased dielectric response [45,46] (Methods). Importantly, under a harmonic approximation for the solid modes, the free energy difference $\Delta F_m$ between the light-matter hybrid and the isolated systems can be understood solely in terms of the frequency-dependent polarizability of the solid, irrespective of the microscopic nature of its collective modes. The model indicates that the free energy of the metallic state $F_m$ is lowered upon lowering the cavity frequency, which is qualitatively consistent with the decrease of the effective critical temperature upon reducing the cavity frequency observed experimentally (Fig. 4A). We stress that the coupling to a single cavity mode quantitatively gives only a non-extensive contribution to the free energy [47]. The total effect on the free energy depends on the phase space of the relevant cavity modes which can be shifted within the solid spectrum. Taking into account this phase space factor, extremely large couplings would be needed for the free energy changes to explain the observed shifts in the transition temperature (see Methods for quantitative estimations).

In the second mechanism presented ù in Fig. 1A, the cavity controls the temperature of the sample and consequently the difference between the external ($T_{ext}$) and the sample temperature ($T_{int}$). The reshaping of the EM density of the states driven by the cavity can indeed induce a change in the emission spectrum of the sample and hence of its temperature [48] (this scenario is represented in Fig. 4D as a renormalization of the temperature axis, blue horizontal arrow). Shorter cavities move the electromagnetic modes to higher frequencies and could effectively decouple the optically active solid modes from the external field, analogously to the Purcell effect [49]. The sample is in thermal contact with the cold finger through the membranes, but it is also in contact with the external photon bath at $T_{ph}$ = 300 K (Fig. 1A, lower right panel). We assume that the thermal transfer from the cold finger to the sample depends only on the difference between $T_{ext}$ and $T_{int}$ through a cavity-independent thermal coupling constant. Conversely, the thermal load on the sample due to the contact with the external photon bath is mediated by the cavity through a coupling constant depending on the cavity geometry (fundamental frequency $\omega_c$ and quality factor $Q$) and on the sample dielectric loss within the employed cavity range. In order to qualitatively illustrate the mechanism, we model the infrared spectrum of 1T-TaS$_2$ as a broad continuum absorption band lying in the GHz range, and exploit the Purcell-based model to extract an effective temperature of the sample $T_{int}(\omega_c, Q)$ depending on the cavity geometry (see Methods for further details). We show that upon increasing the cavity frequency, i.e. by decoupling the electromagnetic active transitions in the sample from the cavity fundamental mode, the cavity induces a cooling of the sample, whose effective temperature ($T_{int}$) reaches the cold-finger one ($T_{ext}$) at high cavity frequencies (Methods). This trend is qualitatively consistent with the effective critical temperature trend observed experimentally with both THz (Fig. 4A) and temperature measurements (Fig. 4B).

In contrast to the first scenario, the phase space restriction is no longer valid in this second mechanism (Fig. 1A), in which an open system is considered and thermal exchanges between material, cold finger and photon bath are allowed. This, together with the experimental evidence that the cavity can coherently modify the sample temperature as a function of the frequency (Fig. 4B), suggests that the Purcell-based mechanism may be the dominant effect in the reported observation. Nevertheless, it is interesting to note that both mechanisms predict the correct dependence of the effective critical temperature on the cavity frequency. This may provide a useful guide for future quantitative theories, which should also take into account the open nature of the system as well as the non-linear interaction between the modes in the solid.

In conclusion, we have demonstrated that the metal-to-insulator transition in 1T-TaS$_2$ embedded in low energy THz cavities can be reversibly controlled by the cavity geometry. The evidence points to a scenario in which the cavity electrodynamics modifies the effective sample temperature. Our results provide a previously unknown control parameter in the rich phase diagram of quantum materials and enable tailoring the equilibrium collective properties in correlated materials by engineering their electromagnetic environment.

# METHODS

## 1. EXPERIMENTAL DETAILS

### 1. EXPERIMENTAL SET-UP

The set-up employed in the experiment [3] is sketched in Extended Data Fig.1A. The cryogenic THz cavity shown in the inset is composed of two cryo-cooled piezo-controlled semi-reflecting mirrors between which the sample is inserted. The movement of both cavity mirrors is ensured by three piezo-actuators (N472-11V, Physik Instrumente) having a travel range of 7 mm and a minimum incremental motion of 50 nm with a designed resolution of 5 nm. The mirrors are mounted on copper holders, and they are cryo-cooled by means of copper braids directly connected to the cold finger of the cryostat. In order to ensure the tuning of the mirrors distance at cryogenic temperatures the piezo actuators are thermally decoupled from the mirrors supports. The thermal decoupling is realized by placing between the piezo actuators and the mirrors holders a PEEK disk, on which the piezoelectrics actually act, and three ceramic cylinders. The chamber shown in Extended Data Fig. 1A (inset) is mounted on a flow cryostat with a temperature feedback circuit enabling temperature scans at fixed cavity length. The temperature is read both on the sample holder and on the mirrors holders. The deviation in temperature is less than 1 K for a measurement at a fixed temperature.

The cavity semi-reflecting mirrors were fabricated by evaporating a thin bilayer of titanium-gold (2-15 nm) on a 2 mm crystalline quartz substrate. We measured the transmission amplitude of a single cavity mirror to be 15 % in the THz spectral range of the experiment with no apparent spectral features.
The approximatively 10-μm thick 1T-$TaS_2$ sample is mounted between the two mirrors in a copper sample holder directly connected to the cold finger of the cryostat and sealed between two silicon nitride ($Si_3N_4$) membranes (grown by low-pressure chemical vapour deposition) with a window size of 11 x 11 $mm^2$ and a 2 μm thickness. The membranes are supported on a 13 x 13 $mm^2$ silicon frame and do not show any spectral dependence in the employed THz spectral range.

We employed broadband THz spectroscopy to characterize the metal-to-insulator transition for different cavity settings. Single-cycle THz pulses are generated via the acceleration of the photoinduced carriers in a large-area GaAs-based photoconductive antenna (PCA). The photoexcitation is achieved by pumping the PCA with an ultrashort laser pulse (50 fs pulse duration, 745 nm central wavelength, 6 μJ/pulse energy) from a commercial 50 kHz pulsed laser + Optical Parametric Amplifier (OPA) system (Pharos + Orpheus-F, Light Conversion). The emitted collimated THz beam is then focused on the sample mounted inside the cavity (inset of Extended Data Fig. 1A). We estimated the THz spot diameter at the focus position to be ~ 1.5 mm, hence smaller than the lateral dimensions of the 1T-$TaS_2$ crystal (4 mm x 4 mm). The field transmitted through the sample is probed by standard Electro-Optical-Sampling (EOS) in a 0.5 mm ZnTe crystal with a weak read-out pulse (50 fs pulse duration, 745 nm central wavelength, < 100 nJ/pulse). We present in Extended Data Fig. 1B the measured electric field of the generated THz pulse and its calculated Fourier spectrum (Extended Data Fig. 1C). The input field is a nearly single-cycle THz pulse with a spectral content up to 6 THz, as highlighted in the logarithmic scale plot in the inset of Extended Data Fig. 1C. We estimated the signal-to-noise of the detected THz field to be 4.6 x $10^4$ and a temporal phase stability ≤ 30 fs.

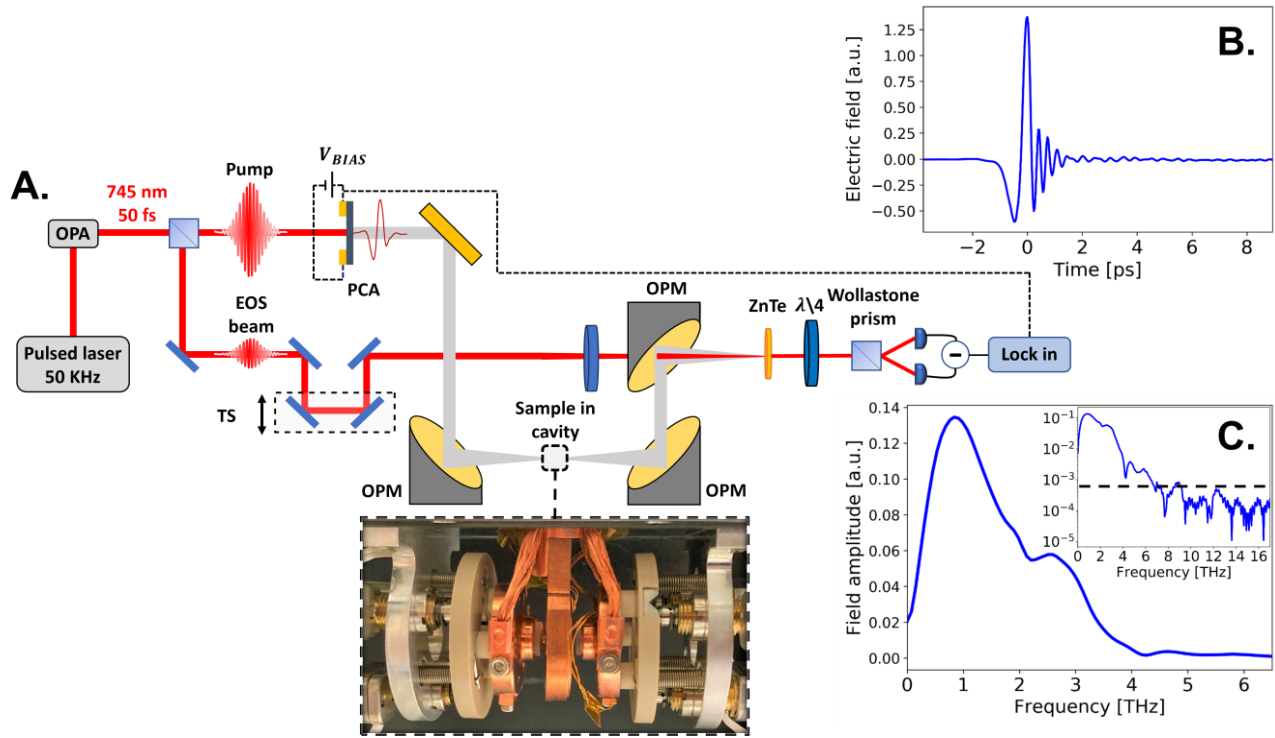

**Extended Data Fig 1**: **Experimental set-up**. **A.** Sketch of the THz time domain spectrometer. In the inset, the photograph of the tunable cryogenic cavity composed of two cryo-cooled moving mirrors within which the sample is embedded. **B.** Free-space nearly single-cycle THz field employed in the experiments detected trough Electro-Optical Sampling (EOS) in a 0.5 mm ZnTe crystal. **C.** Fourier transform of the nearly single-cycle THz field in free space. In the inset, the Fourier spectrum is plotted in logarithmic scale to highlight the spectral content of the THz field up to approximatively 6 THz. The black dashed line in the logarithmic plot indicates the noise level.

## 2. SAMPLE PREPARATION

The single crystals from which flakes are exfoliated are grown from the elements (purity of Ta: 99.95% and S: 99.999%) by the vapor phase transport method at 850 °C with 1.5 mg/cm$^3$ excess S, quenched from the growth temperature to a room temperature water bath to retain the 1T polytype. Single crystal XRD confirms that the pure single 1T phase is retained after the quench (a = b = 3.357, c = 5.91), with the Ta:S composition ratio determined by EDS to be 33:66 +/-1 atomic %. STM measurements show rare-earth impurity on the surface layer content consistent with the impurity concentrations given by the supplier (Alfa Aesar, Merck).

# 2. MEASUREMENT PROTOCOLS AND DATA ANALYSIS

## 1. CHARACTERIZATION OF THE EMPTY CAVITY

To estimate the quality factor of the cavity, we characterized the response of the empty cavity, i.e. when the THz field passes only through the silicon nitride membranes within the mirrors. The quality factor quantifies the photon lifetime inside the cavity and, subsequently, the coupling strength between the cavity mode and the targeted material excitation, which is ruled by the bare cavity dissipations.

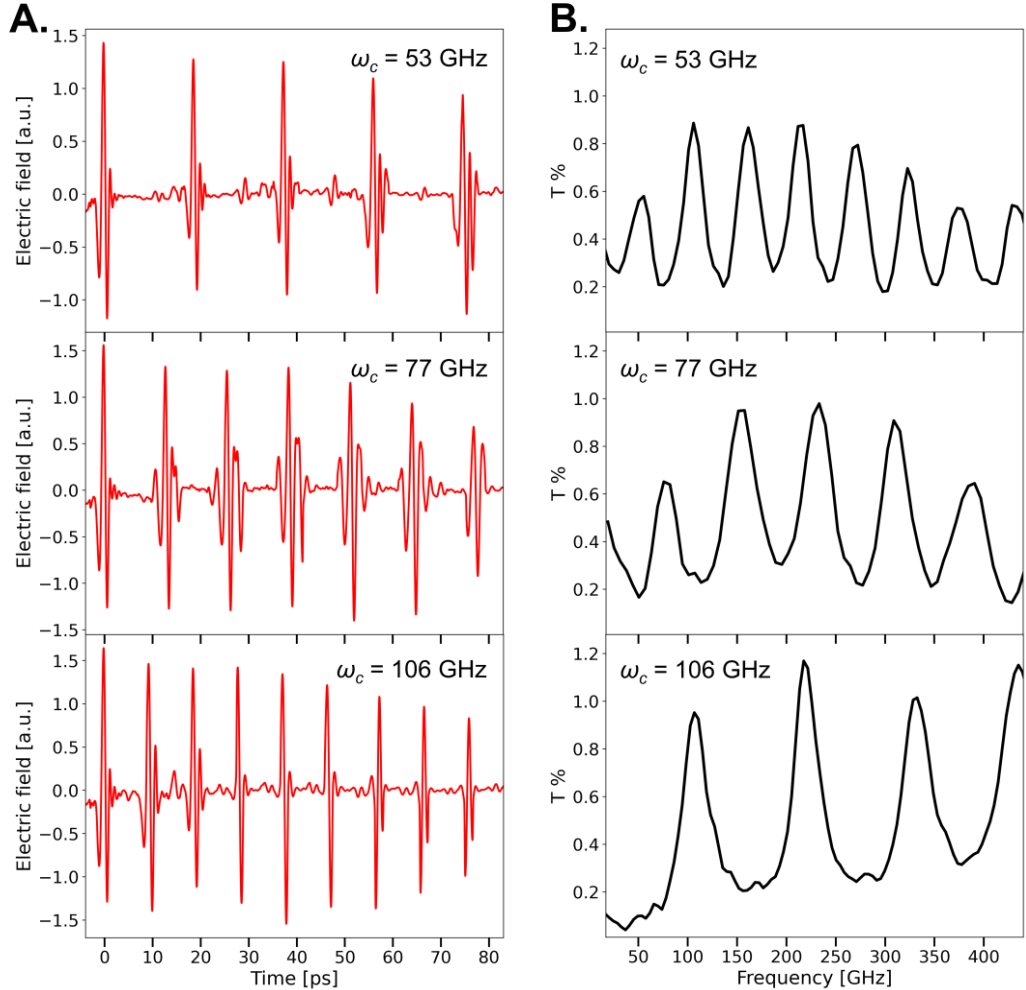

**Extended Data Fig 2**: **THz characterization of the empty cavity**. **A.** Time domain THz fields measured at the output of the empty cavity for three representative cavity frequencies $\omega_c$ indicated in legend. **B.** Cavity transmission spectra calculated from the fields shown in A. proving the tunability of the cavity fundamental mode.

The two cavity mirrors were set parallel to each other and perpendicular to the THz beam by aligning the multiple reflection of the pump optical beam, which propagates collinearly with the THz beam.
In Extended Data Fig. 2 we plot the time domain THz field transmitted through the Fabry-Pérot cavity (Extended Data Fig. 2A) and the corresponding spectral content (Extended Data Fig. 2B) for three representative values of the cavity fundamental mode $\omega_c$ among the one used in the experiment ($\omega_c$ = 53, 77, 106 GHz). The cavity transmission spectra plotted in Extended Data Fig. 2B were obtained by taking the ratio between the Fourier spectrum of the time domain traces shown in Extended Data Fig. 2A and the reference free space spectrum presented in Extended Data Fig. 1D. The cavity transmission spectra exhibit the interference Fabry-Pérot modes with a tunable fundamental frequency set by the cavity length.

For the three representative cavities shown in Extended Data Fig. 2 we estimated the quality factors $Q$ to be 3.3, 3.6 and 3.5 for the 53, 77, and 106 GHz cavity, respectively. The latter were calculated as the ratio between the fundamental cavity frequency and its bandwidth defined as the full width half maximum of the transmission peak of the fundamental mode.
This estimation proves that, for all the cavity lengths that we studied, the bare quality factor (and hence the incoherent photon losses) can be considered independent on the cavity frequency.

## 2. CHARACTERIZATION OF THE SAMPLE THERMALIZATION TIME

In order to prove that all the measurements have been performed in a stationary regime, we estimated the thermalization time of the sample by delaying the THz acquisition by different amounts of time. As shown in Extended Data Fig.3, no significant variation in the effective critical temperature and in the slope of the phase transition occurs for $\Delta t \geq 2$ min. Since all the measured THz traces in the experiment are the result of a 20 minute integration time at each temperature step with a 5 minutes waiting time before the first THz acquisition. Therefore, we can safely rule out that the observed inhomogeneous-like feature of the phase transition is due to a measurement waiting time less than the sample thermalization and it can be instead likely ascribed to intrinsic inhomogeneities of the sample, which can smear out the phase transition [43]. Another factor which could be responsible for the smearing out of the metal-to-insulator transition in single crystal 1T-TaS$_2$ samples is substrate strain. Strain plays indeed a large role in the metal-to-insulator transition and can shift and broader the transition temperature substantially [44]. Additionally, when mounted on membranes, homogeneity of the temperature, in particular the in-plane one, may also broaden the transition.

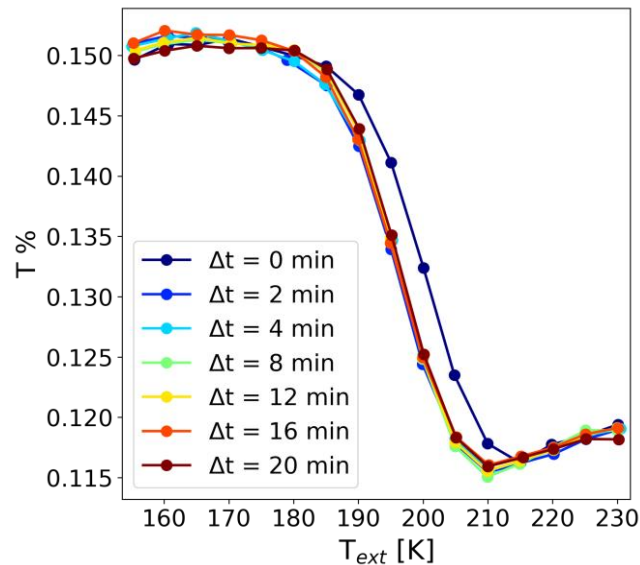

**Extended Data Fig. 3: Dependence of the observed metal-to-insulator phase transition on the waiting time.** The temperature evolution of the low frequency transmission (0.2 THz < ω < 1.5 THz) is plotted for different waiting times before starting the THz acquisition.

# 3. DETERMINATION OF THE EFFECTIVE CRITICAL TEMPERATURE

In this section we present the method used to extract from the THz transmission data the effective critical temperatures discussed in the main manuscript.

We show in Extended Data Fig.4A the temperature evolution of the integrated low frequency transmission (0.2 THz < ω < 1.5 THz) associated to the onset of the metallicity. Extended Data Fig. 4B shows instead the temperature evolution of the transmitted spectral weight around the 1.58 THz phonon integrated in the range 1.53 THz – 1.62 THz. This integration range corresponds to the phonon bandwidth.

In order to estimate the effective critical temperature at each cavity length, we interpolated the metallic and phononic temperature response and calculated the derivative of the interpolated curve. The obtained temperature-derivative for the free space sample is presented in the lower panels of Extended Data Fig. 4A and Extended Data Fig. 4B for the metallic and phononic response, respectively.

We set the effective critical temperature of the phase transition ($T_c^{eff}$) to be the maximum of the derivative of the interpolated curve. The error associated to each $T_c^{eff}$ is the standard deviation of the effective critical temperatures estimated for three consecutive scans. The robustness of this procedure is validated by the fact that the effective critical temperatures estimated from the metallic response are compatible with the ones estimated from the phonon spectral response.

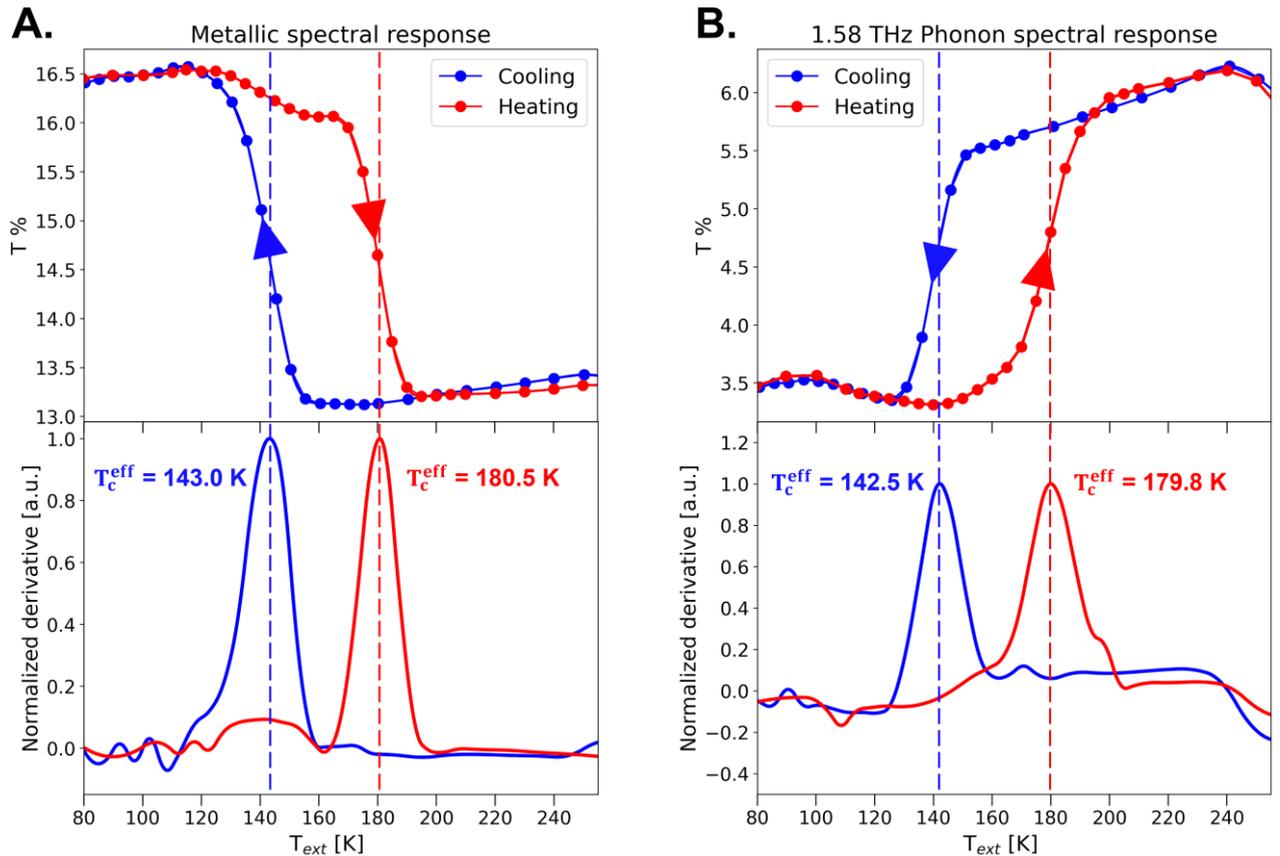

**Extended Data Fig 4: Determination procedure of the effective critical temperature of the metal-to-insulator transition.** **A.** In the top panel the temperature evolution of the integrated low frequency transmission (0.2 THz < ω < 1.5 THz integration range) upon heating and cooling (circled markers). The solid line is the result of an interpolation. In the lower panel the derivative of the interpolated curve whose maximum sets the effective critical temperature of the phase transition. **B.** In the top panel the temperature evolution of the integrated 1.58 THz phonon transmission (1.53 THz < ω < 1.62 THz integration range) and its interpolation. In the lower panel the derivative of the interpolated phonon response across the phase transition.

# 3. THEORETICAL MODELS

## 1. FREE ENERGY PICTURE

In this section we introduce the phenomenological model providing a qualitative estimation of the renormalization of the free energy of the metallic NC state due to cavity electrodynamics.

Let us consider a solid with given dielectric properties, characterized by the polarizability $\alpha(\omega)$ which determines the response of the transverse polarization density to the electric field, $\vec{P}(\omega) = \epsilon_0 \alpha(\omega) \vec{E}(\omega)$. The polarizability $\alpha(\omega)$ is related to the relative dielectric function $\epsilon(\omega)$ as $\epsilon(\omega) = 1 + \alpha(\omega)$. A non-zero polarizability implies that there are modes in the solid which can hybridize with the electromagnetic field, which in turn leads to a change of the free energy when the system is put in the cavity. In order to understand the effect of the cavity on the free energy of the system, we evaluate the difference

$$\Delta F = F_{tot} - F_{mat} - F_{cav} \tag{1}$$

between the total free energy of the coupled system ($F_{tot}$) and the free energies of the uncoupled solid ($F_{mat}$) and of the electromagnetic field ($F_{cav}$). A key observation is that as long as the solid is approximately described by a harmonic theory, $\Delta F$ can be determined from the knowledge of the experimentally accessible dielectric function alone, independent of microscopic details such as the precise nature of the electromagnetically active modes. In short, the reason is that in a harmonic theory one can exactly integrate out the modes of the solid, so that the resulting effective action of the cavity, which then determines $\Delta F$, is given in terms of the linear response functions of matter.

We will make a further simplification in line with the present experimental setting, and assume that the volume $V_m$ of the solid is small compared to the cavity volume $V$, $V_m/V \ll 1$. This approximation is valid for the experimental setting since the cavities employed have fundamental frequencies in the low THz region, while the sample thickness is ~ 10 μm. With this approximation, as we will show below, for a single cavity mode with fundamental frequency $\omega_c$ the free energy renormalization $\Delta F$ due to the light-matter coupling (Eq. 1) is given by $\Delta F(\omega_c, T) = \frac{V_m}{V} f(\omega_c, T)$, where

$$f(\omega_c, T) = \frac{1}{\pi} \int_0^\infty d\omega\, \alpha''(\omega)\omega \frac{b(\omega_c, T)\omega_c - b(\omega, T)\omega}{\omega_c^2 - \omega^2}. \tag{2}$$

In the previous equation $b(\omega, T) = (e^{\omega/T} - 1)^{-1}$ is the Bose function and $\alpha''(\omega)$ the imaginary part of the solid polarizability (dielectric loss).

The total free energy change $\Delta F$ is a thermodynamically extensive quantity, which arises from the coupling to a continuum of cavity modes with transverse momentum $q$ and a discrete mode index $n$ (Extended Data Fig. 5A). For simplicity, instead of summing Eq. 2 over all cavity modes $\omega_c \equiv \omega_{q,n}$, we will first analyze the single mode result (Eq. 2) for the lowest cavity frequency ($\omega_c = \pi c/L$, with $L$ the cavity length) to understand the qualitative functional dependence of $\Delta F$ on the temperature and on the cavity parameters. In order to estimate the order of magnitude of the total effect of all modes, the result will then be multiplied with a phase space factor that counts the number of modes $N_{mode}$ that are affected by the cavity.

To analyze the free energy renormalization (Eq. 2), we assume that the solid polarizability $\alpha(\omega)$ gives rise to a broad continuum absorption band that can be fitted by the response of a strongly damped oscillator:

$$\alpha(\omega) = \alpha(0) \frac{\Omega^2}{\Omega^2 - \omega^2 - i\omega\gamma}. \tag{3}$$

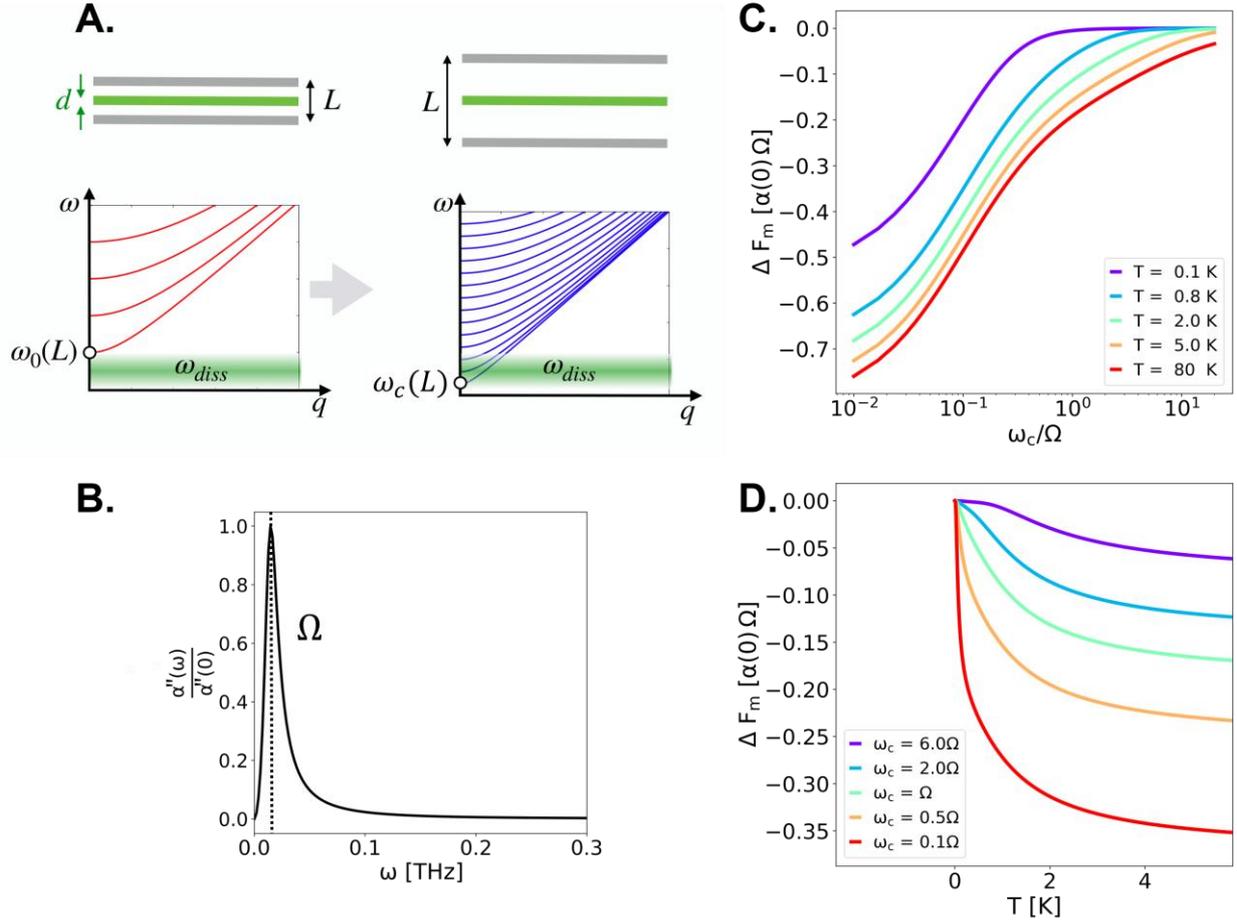

**Extended Data Fig 5: Cavity-induced renormalization of the free energy of the metallic phase. A.** Free energy model setting. Upper panels: coplanar cavity with a thin slab of matter (thickness $d$) inside a cavity of length $L$. Lower panels: Sketch of the cavity modes dispersion and of the absorption solid band (green shaded region centered at $\omega_{diss}$). As $L$ is increased, modes are pulled inside and below the absorption band of the solid. The cavity fundamental mode is indicated with $\omega_c(L)$. **B.** Dielectric loss spectrum $\alpha''(\omega)$ ($\Omega$ = 15 GHz, $\gamma$ = 20 GHz) employed for the calculations. The spectrum has been normalized by the static contribution to the polarizability $\alpha(0)$. **C.** Renormalization of the metallic free energy $\Delta F_m$ as a function of the cavity frequency for different temperatures. The cavity frequencies $\omega_c$ are normalized by $\Omega$ = 15 GHz. D. Renormalization of the metallic free energy $\Delta F_m$ as a function of the temperature for different cavity frequencies above and below resonance $\omega_c = \Omega$.

Here $\Omega$ corresponds to the central frequency of the material's mode, $\gamma$ is the linewidth, and $\alpha(0)$ is the contribution of the mode to the static polarizability. The latter also measures the total spectral weight in the absorption band and therefore serves as a phenomenological measure of the effective coupling strength. The dielectric loss $\alpha''(\omega)$ adopted for the estimations is presented in Extended Data Fig. 5B. We set a central frequency $\Omega$ = 15 GHz and a frequency damping $\gamma$ = 20 GHz so that no significant contribution to the solid dielectric loss is present in the THz region ($\omega > 0.1$ THz).

We show in Extended Data Fig. 5C the dependence of the free energy renormalization of the metallic phase (Eq. 2) as a function of the cavity frequency $\omega_c$ when the latter is swept through the mode centered at $\Omega$. The model indicates that the free energy of the metallic state is lowered upon lowering the cavity frequency, which is qualitatively consistent with the decrease of the effective critical temperature upon reducing the cavity frequency observed experimentally. The renormalization of the metallic free energy is larger for larger temperatures, indicating that it is related to the thermal population of the low energy mode. We stress that the temperature in the experiment is well above $\Omega$. Extended Data Fig. 5D shows the free energy renormalization as a function of temperature for different cavity frequencies $\omega_c$ below and above the resonance $\omega_c = \Omega$. The free energy of the metallic phase is lowered and becomes steeper when the cavity frequency is lowered (i.e. opening the cavity). This trend is consistent with the interpretation of the experimental observation in the main text highlighted in Fig. 4D.

It should be stressed, however, that the absolute changes of the total free energy are expected to be rather small. As mentioned above, the single mode result $\Delta F(\omega_c, T) = \frac{V_m}{V} f(\omega_c, T)$ should be integrated over all modes or, for a simple estimate, multiplied with a phase space factor $N_{modes}$. If the latter is simply taken to account for all modes below a certain cutoff $\omega_{cut}$ in a volume $V$, we have $N_{modes} = V/\lambda_{cut}^3$, up to constants of order one. Hence the free energy change $N_{modes} \times \frac{V_m}{V} f(\omega_c, T)$ per volume $V_m$ is given by the amount $f(\omega_c, T)$ per volume $\lambda_{cut}^3$. The changes of $f(\omega_c, T)$ upon modifying the cavity frequency are of the order of $\alpha(0)\Omega$ (see Extended Data Fig. 5C, D), thus corresponding to an energy density $\alpha(0)\Omega/\lambda_{cut}^3$. This value has to be compared with the condensation energy density of the phase transition, which is $Q \approx 6 J/mm^3 \approx 3.6 \times 10^{10} eV/\mu m^3$ [50]. With $\Omega$ in the sub-meV range, very large couplings $\alpha(0)$ would be needed, even with a cutoff $\lambda_{cut}$ in the optical range (which is clearly an upper bound, as optical frequencies are hardly affected by the present cavity setting).

We therefore conclude that although the free energy renormalization in the cavity $\Delta F$ follows the correct trend (lowering the free energy of the nearly commensurate phase as the cavity is opened), it is not sufficient to explain the experimental observation. While it will certainly be also interesting to investigate future theoretical interpretations which go beyond the harmonic theory, this puts more emphasis on the second mechanism (Purcell-like effect) discussed in the present manuscript.

Finally, let us conclude with the derivation of Eq. 2. Let us start from a general harmonic model in which one mode of the electromagnetic field couples to a continuum of modes in the solid. The Hamiltonian is a general Dicke-type Hamiltonian:

$$H = \frac{1}{2}\sum_a \left( \Omega_a^2 \left[p_a + \frac{\kappa \gamma_a}{\Omega_a^2}\Pi\right]^2 + x_a^2 \right) + \frac{1}{2}(\Pi^2 \omega_q^2 + X^2), \tag{4}$$

where $X$ and $\Pi$ are the canonical quadratures of the electromagnetic field, and $p_a$ and $x_a$ the quadratures of the modes in the solid. The solid normal modes have frequencies $\Omega_a$ in the absence of the coupling to the cavity field. The electromagnetic field quadrature $\Pi$ is related to the vector potential (which is spatially homogeneous throughout the solid) by the relation

$$\hat{A} = \hat{n}\sqrt{\frac{1}{\epsilon_0 V}}\Pi, \tag{5}$$

where we have indicated with $V$ the cavity mode volume, and with $\hat{n}$ the polarization direction. The electric field is connected to the vector potential by the temporal derivative $E = -\dot{A}$. The complete square light-matter coupling in Eq. 4, with coupling constants $\gamma_a$, corresponds to a minimal coupling of the modes in the solid to the vector potential. The coupling between the solid oscillators and the single cavity mode scales as $\kappa \gamma_a$ where $\kappa^2 = V_m/V$ is the volume fraction of the cavity filled with the solid. Note that, as mentioned above, the final result for $\Delta F$ will be expressed in terms of the polarizability, so that the detailed choice of the parameters $\Omega_a$ and $\gamma_a$ does not enter.

The aim is now to calculate the free energy difference $\Delta F$ within this model. Denoting by $\eta_\alpha$ and $\eta_\alpha^{(0)}$ the normal modes energies of the coupled and uncoupled system, respectively, the free energy difference (Eq. 1) is simply given by

$$\Delta F = \frac{1}{\beta}\sum_\alpha \left[ ln(1 - e^{-\beta \eta_\alpha}) - ln\left(1 - e^{-\beta \eta_\alpha^{(0)}}\right) \right], \tag{6}$$

where $ln(1 - e^{-\beta\eta})/\beta$ is the free energy of an oscillator with frequency $\eta$. The $\eta_\alpha^2$ are given by the eigenvalues of the dynamical matrix $\mathcal{D}$ corresponding to the Hamiltonian (Eq. 4),

$$\mathcal{D} = \begin{pmatrix} \widetilde{\omega}_c^2 & \kappa\gamma_1 & \kappa\gamma_2 & \cdots \\ \kappa\gamma_1^* & \Omega_1^2 & 0 & \\ \kappa\gamma_2^* & 0 & \Omega_2^2 & \\ \vdots & & & \ddots \end{pmatrix}. \tag{7}$$

Here $\widetilde{\omega}_c^2$ is the shifted cavity frequency $\widetilde{\omega}_c^2 = \omega_c^2 + \kappa^2 \sum_a \frac{|\gamma_a^2|}{\Omega_a^2}$ as a consequence of the coupling with the solid degrees of freedom. We determine these normal modes perturbatively in the solid volume fraction $\kappa^2 \ll 1$. The perturbative expansion for the cavity mode (entry 0 in Eq. 7) reads

$$\eta_0^2 = \omega_c^2 + \kappa^2 \sum_a \frac{|\gamma_a^2|}{\Omega_a^2} + \kappa^2 \sum_a \frac{|\gamma_a^2|}{\omega_c^2 - \Omega_a^2}, \tag{8}$$

while for the matter modes we have:

$$\eta_a^2 = \Omega_a^2 + \kappa^2 \frac{|\gamma_a^2|}{\Omega_a^2 - \omega_c^2}. \tag{9}$$

One can then linearize Eq. 6 in $\delta\eta$, leading to $\Delta F = \kappa^2 \sum_\alpha b(\eta_\alpha^{(0)})\delta\eta_\alpha$, and insert the perturbative expressions of the cavity (Eq. 8) and solid (Eq. 9) eigenmodes. With some straightforward manipulations, this gives the result of Eq. 2, with the function $\alpha''(\omega)$ of the form

$$\alpha''(\omega) = \pi \sum_a \frac{|\gamma_a|^2}{2\Omega_a^3} [\delta(\omega - \Omega_a) - \delta(\omega + \Omega_a)]. \tag{10}$$

Finally, we need to confirm that this expression (Eq. 10) is precisely the imaginary part of the polarizability within the model of Eq. 4. A simple link is made via the dielectric loss. When the system is driven with a time-dependent field, the absorbed energy per volume is the time-average of $\vec{E}(t)\partial_t \vec{P}(t)$. With the above definition of the polarizability, the loss under a field $A(t) = \hat{n}A_\omega e^{-i\omega t} + h.c.$ is

$$\Gamma(\omega) = 2\omega^3 \epsilon_0 \alpha''(\omega)|A_\omega|^2. \tag{11}$$

On the other hand, in the model (Eq. 4) we can calculate the energy absorption due to a time dependent classical vector potential, which by means of Eq. 5 is introduced by replacing $\Pi \to \Pi + \sqrt{V\epsilon_0}A(t)$. Fermi's golden rule (or equivalently the Kubo linear response formalism) gives

$$\Gamma_A(\omega) = |A_\omega|^2 2\omega \sum_a |\gamma_a|^2 \chi_{aa}''(\omega)\epsilon_0 V\kappa^2 = V_m|A_\omega|^2 2\omega \sum_a |\gamma_a|^2 \chi_{aa}''(\omega)\epsilon_0, \tag{12}$$

with the spectral function

$$\chi_{aa}''(\omega) = \frac{\pi}{2\Omega_a}[\delta(\omega - \Omega_a) - \delta(\omega + \Omega_a)] \tag{13}$$

of the single mode $a$. Comparing the two expressions in Eq. 11 and Eq. 12 shows that Eq. 10 is the result for the polarizability.

## 2. CONTROL OF DISSIPATIONS THROUGH CAVITY ELECTRODYNAMICS

In this section we discuss the Purcell-like scenario mentioned in the main manuscript, i.e. the mechanism in which the observed changes in the effective critical temperature could be related to a cavity control of the dissipations, analogously to the Purcell effect. In this scenario, the reshaping of the electromagnetic density of states at the sample position due to the cavity electrodynamics could result in a modification of the sample's thermal load and subsequently of its temperature.

In order to estimate this effect, we proceed as indicated in Extended Data Fig. 6A. The sample is in thermal contact with the cold finger through the membranes, but it is also in thermal contact with the external photon bath at $T_{ph}$ = 300 K. We assume that the thermal transfer from the cold finger to the sample depends only on the difference between the cold-finger temperature ($T_{ext}$) and the sample effective temperature ($T_{int}$). Conversely, we assume that the thermal load on the sample due to the contact with the external photon bath is mediated by the cavity, in analogy with the Purcell effect. Under these hypotheses, we can write two rate equations describing respectively the cavity-independent heat flow between the cold finger and the sample:

$$Q_{ext-int} = K_{ext-int}\,(T_{ext} - T_{int}), \tag{14}$$

and the cavity-mediated heat transfer between the sample and the external photon bath:

$$Q_{ph-int}(\omega_c, Q) = K_{ph-int}(\omega_c, Q)(T_{ph} - T_{int}). \tag{15}$$

In the previous equations $K_{ext-int}$ represents the cavity-independent coupling constant between the cold finger and the sample, while $K_{ph-int}(\omega_c, Q)$ the coupling constant between the sample and the photon bath, which depends on the cavity geometry, i.e. on the fundamental frequency $\omega_c$ and on the quality factor $Q$.

The coupling constant $K_{ph-int}(\omega_c, Q)$ between the sample and the photon bath can be expressed as the joint density of states of the solid $\rho_{Solid}(\omega)$ and of the cavity $\rho_{Cavity}(\omega_c, Q)(\omega)$, with the latter multiplied by the Boltzmann distribution at the photon bath temperature $T_{ph}$ = 300 K:

$$K_{ph-int}(\omega_c, Q) = \int_0^\infty d\omega\, \rho_{Cavity}(\omega_c, Q)(\omega)\, \rho_{Solid}(\omega)\, e^{-\frac{\omega}{K_B T_{ph}}}. \tag{16}$$

Considering a continuum broad mode centered at $\Omega$ = 15 GHz and with a spectral linewidth $\gamma$ = 20 GHz (as for the free energy model described in the previous section), the solid density of states associated to the material's excitations can be expressed through the dielectric loss per unit frequency as:

$$\rho_{Solid}(\omega) = \frac{\alpha''(\omega)}{\Omega} = \alpha(0)\frac{\Omega\gamma\omega}{(\omega^2 - \Omega^2)^2 + (\gamma\omega)^2}. \tag{17}$$

Conversely, the multimode cavity density of states takes the form:

$$\rho_{Cavity}(\omega_c, Q)(\omega) = \sum_{n=0}^{+\infty} \frac{\gamma_{cav}}{(\omega - n\omega_c)^2 + (\gamma_{cav})^2}, \tag{18}$$

where $\gamma_{cav}$ represents the linewidth of the bare cavity which is related to the quality factor $Q$ by the relation $Q = \omega_c/\gamma_{cav}$. The quality factor of the empty cavity is set by the experimental conditions (see Section "Characterization of the empty cavity" of the Methods).

In Extended Data Fig. 6B we present a plot of the solid density of state and of the cavity density of states multiplied by the Boltzmann distribution at the photon bath temperature $T_{ph}$ = 300 K.

Under stationary conditions, the thermal flow from the cold finger to the sample $Q_{ext-int}$ equals the cavity-mediated heat transfer between the sample and the photon bath $Q_{ph-int}(\omega_c, Q)$, that is $Q_{ext-int} + Q_{ph-int}(\omega_c, Q) = 0$. At equilibrium, we can subsequently calculate an effective sample temperature $T_{int}(\omega_c, Q)$, which takes the form:

$$T_{int}(\omega_c, Q) = \frac{K_{ph-int}(\omega_c, Q)T_{ph} + K_{ext-int}T_{ext}}{K_{ph-int}(\omega_c, Q) + K_{ext-int}}. \tag{19}$$

The temperature ratio between the sample and the cold finger $\frac{T_{int}(\omega_c, Q)}{T_{ext}}$ as a function of the cavity fundamental frequency is plotted in Extended Data Fig. 6C for different cold-finger temperatures. We stress that the renormalization of the sample effective temperature scales with the cavity-solid joint density of states $K_{ph-int}(\omega_c, Q)$, and hence with the total spectral weigth within the solid absorption band $\alpha(0)$. A larger renormalization of the sample's temperature is hence expected for a larger oscillator strength of the solid's modes.

While the density of states of the electromagnetic field can be enhanced inside the optical cavity with respect to free space, potentially enhancing radiative transitions in materials in resonance with the cavity, shorter cavities move the electromagnetic modes to higher frequencies and could effectively decouple the optically active material's modes from the external field, analogously to the Purcell effect.

We note that, upon increasing the cavity frequency, i.e. by reducing the coupling of the EM active modes from the cavity fundamental mode, the model predicts a decrease of the temperature ratio $T_{int}(\omega_c, Q)/T_{ext}$, consistent with a decrease of the effective temperature of the sample (Fig. 4B). This trend is qualitatively consistent with what observed experimentally with THz spectroscopy (Fig 4A), i.e. an increase of the effective critical temperature ($T_c^{eff}$) upon increasing the cavity frequency.

Moreover, upon lowering the cavity frequency, the coupling between the active optical transitions in the material and light can be enhanced within a frequency range, and the cavity may effectively enhance the absorption of the external blackbody radiation from the sample [51], heating up the sample with respect to free space conditions. Our measurements indicate that the cavity frequency range explored in the experiments is at higher frequencies with respect to the relevant absorption modes in the solid.

As highlighted in Extended Data Fig. 6C, the cavity-mediated modification of the sample-photon bath dissipations is more efficient at lower cold-finger temperatures, i.e. when the difference between the temperature of the photon thermal bath ($T_{ph}$) and $T_{ext}$ is larger. Since the phase transition upon cooling 1T-TaS$_2$ occurs at a lower temperature with respect to the phase transition upon heating it, we expect the Purcell-like effect to be more efficient in shifting the apparent cooling critical temperature with respect to the heating one. This prediction could justify the effective shrinking of the hysteresis observed by sweeping the cavity mode from lower towards higher frequencies (Fig. 4C).

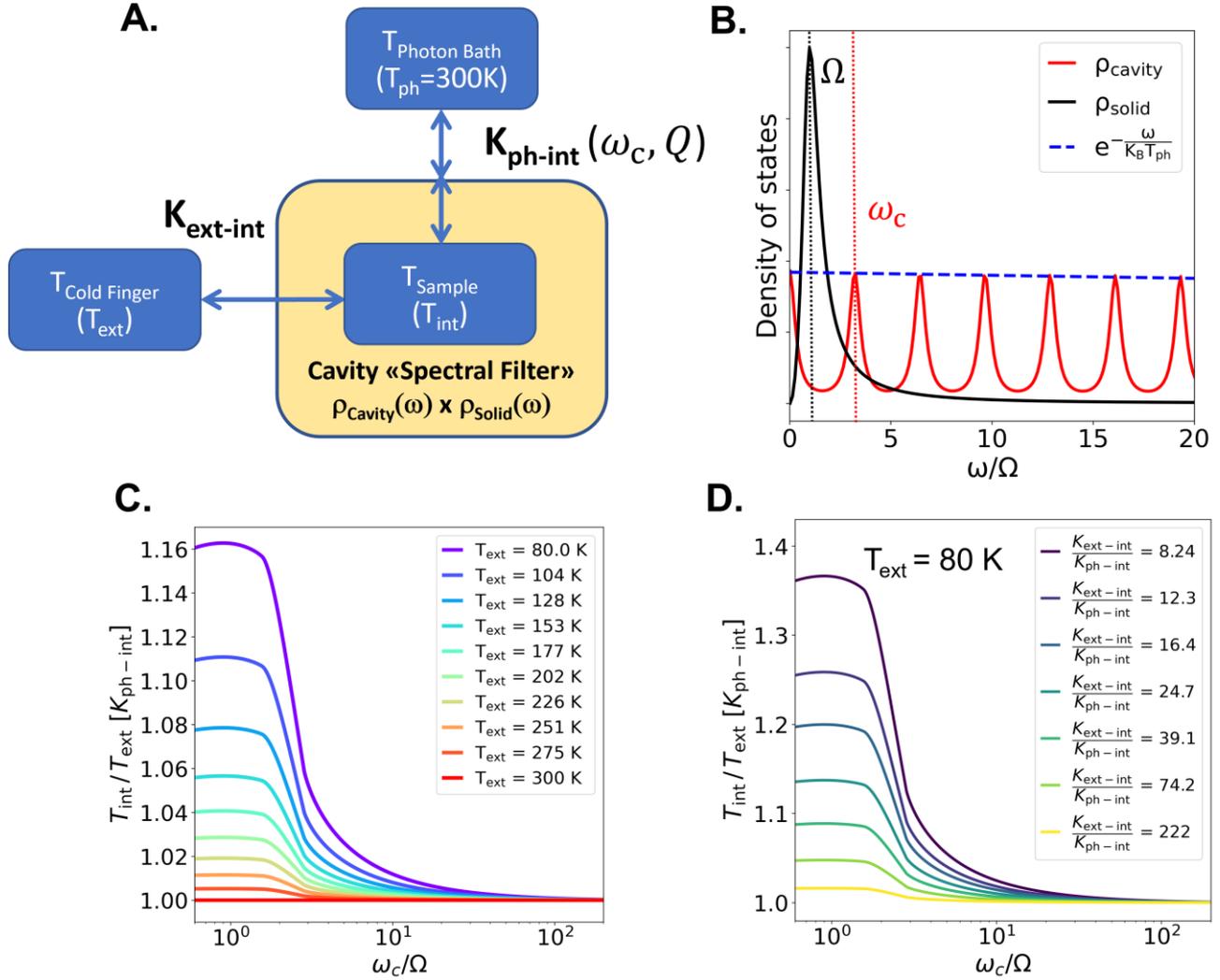

**Extended Data Fig 6: Cavity control of sample dissipations. A.** Schematic representation of the thermal loads on the sample determined by its coupling with the cold finger through the cavity-independent factor $K_{ext-int}$ and with the photon thermal bath through the cavity-dependent factor $K_{ph-int}(\omega_c, Q)$. **B.** Density of states of the solid (peaked at the mode frequency $\Omega$) and of the cavity (peaked at multiples of the fundamental mode $\omega_c$). The cavity density of states is multiplied by the Boltzmann distribution at the temperature of the photon bath $T_{ph}$ = 300 K. **C.** Dependence of the temperature ratio $T_{int}(\omega_c, Q)/T_{ext}$ as a function of the cavity frequency for different temperatures of the cold finger $T_{ext}$. The absolute temperature renormalization scales with $K_{ph-int}(\omega_c, Q)$. **D.** Evolution of the temperature ratio $T_{int}(\omega_c, Q)/T_{ext}$ upon tuning the cavity frequency for different values of the cavity-independent coupling constant $K_{ext-int}$ at a fixed cold-finger temperature $T_{ext}$ = 80 K. The values of the cavity-independent constant $K_{ext-int}$ indicated in the legend have been normalized by $K_{ph-int}(\omega_c, Q)$ evaluated at $\omega_c = \Omega$.

In Extended Data Fig. 6D we prove that the trend presented in Extended Data Fig. 6C is qualitatively independent on the thermal coupling constant between the sample and the cold finger $K_{ext-int}$. A change in $K_{ext-int}$ in the employed cavity frequency range acts only as a scaling factor of the cavity frequency trend. The results shown in Extended Data Fig. 6D has been calculated for a representative cold-finger temperature $T_{ext}$ = 80 K.

Lastly, we point out that the renormalization of the sample effective temperature induced by the cavity is more efficient when the thermal coupling between the sample and the cold finger is smaller. At very high thermal couplings $K_{ext-int}$ we expect indeed the contribution to the sample's temperature of the cavity-dependent interaction with the photon bath, and hence the renormalization of $T_{int}$, to be negligible.

Further studies are needed to provide a quantitative estimate of the cavity dependent total radiative heat load experienced by the sample in the optical cavity. An increased sensitivity in this respect could be provided by cavity design featuring a better thermal isolation between the sample and the cold finger.

## 4. TEMPERATURE MEASUREMENTS WITHIN THE CAVITY

We stressed throughout the manuscript that the temperature indicated in all the reported measurements is the cold-finger readout. When performing THZ optical measurements, this choice is mandatory because any thermosensitive device introduced in the cavity would not only impede the THz transmission, but also modify the sample environment. On the other hand, measuring the actual temperature of the sample is crucial to discriminate between the two cavity-mediated scenarios that we proposed (see Section "Theoretical models" of the Methods).

To this aim, we directly measured the temperature – both of the membrane and the sample – in the cavity by sealing of a home-made 20 μm Cr-Al junction within the membranes. In Extended Data Fig. 7 we show a picture of the thermocouple arrangement within the sample mount. Importantly, in order to not have offsets in the temperature readout, all the wires connecting the junction to the output of the cryostat's head were made of Cr and Al of ~ 120 μm. The only discontinuity point is represented by the gold male-female connectors at the output of the sample holder which, as we verified, give no temperature discrepancy.

We highlight that, in this experimental setting, the THz optical measurements cannot be performed; it is therefore not possible to monitor the THz response of the sample as function of its actual (measured) temperature. All the temperature measurements discussed below must be then considered a separate characterization of the temperature of the sample in a cavity geometry which is nevertheless identical to the one used in all the THz measurements discussed in the main manuscript.

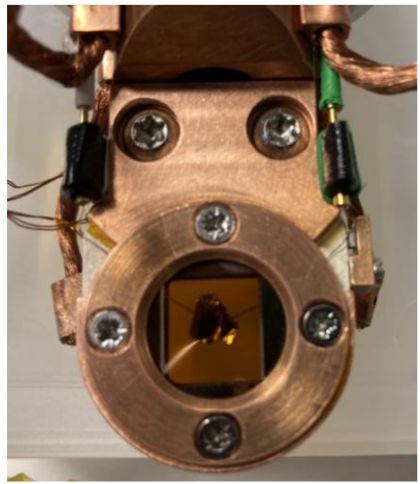

**Extended Data Fig 7: Temperature measurements set-up.** Photograph of the micrometric Cr-Al junction sealed within the membranes and in thermal contact with the sample.

# 5. FINITE-ELEMENTS SIMULATIONS OF INCOHERENT HEATING

To estimate the effect of the incoherent thermal radiation within the cavity, we performed finite-elements simulations exploiting the COMSOL MULTIPHYSICS software. By simulating the incoherent thermal load at the membrane position, we gained insight on the membrane's thermal profile for different cavity configurations.

Let us model the membrane as a grey body having emissivity ε, reflectivity ρ, absorptivity α, and temperature T, and let us assume the incoherent radiative properties of the membrane to be fully described by these four parameters ε, ρ, α, T. The net inward heat flux $Q$ at certain point $x$ on the membrane's surface will be given by the difference between the total arriving radiative flux $G$ (irradiation) and the total outgoing radiative flux $J$ (radiosity):

$$Q(x) = G(x) - J(x). \tag{20}$$

The radiosity $J$ is the sum of the reflected and emitted radiation from the membrane and can be described through the Stefan-Boltzmann equation as:

$$J(x) = \rho G(x) + \epsilon \sigma T^4. \tag{21}$$

By imposing now that the membrane is in thermodynamical equilibrium with the surroundings, i.e. the emissivity ε is equal to the absorptivity α, we can rewrite the reflectivity ρ as:

$$\alpha = \epsilon = 1 - \rho. \tag{22}$$

Thus, the net inward radiative flux of the membrane can be expressed only as a function of $G$, ε, and T as:

$$Q(x) = \epsilon(G(x) - \sigma T^4). \tag{23}$$

Equation 23 has been used in COMSOL as radiation boundary condition for the membrane's surface. The total surface radiation $G$ includes radiation from both the ambient surroundings and from other surfaces. A generalized equation for the irradiative flux is:

$$G(x) = G_m(x) + F_{amb}(x)\, \sigma\, T^4_{amb}, \tag{24}$$

where $G_m$ is the mutual irradiation arriving from other surfaces in the modelled geometry, $T_{amb}$ = 300 K is the temperature of the surrounding environment schematized as a radiative black body, and $F_{amb}$ is the ambient view factor. The latter parameter describes the portion of the view from each point that is covered by ambient conditions. Conversely, $G_m$ is determined by the geometry and the local temperatures of the surrounding surface boundaries. Including the expression of the irradiation $G$ inside Eq. 23, the general expression of the net radiative load at the specific point $x$ on the membrane is:

$$Q(x) = \epsilon(G_m(x) + F_{amb}(x)\sigma\, T^4_{amb} - \sigma T^4). \tag{25}$$

This equation has been used by COMSOL to compute the net radiative transfer at each point $x$ of the membrane's surface. We stress that Eq. 25 results in a linear equation system in $Q(x)$ that must be solved in parallel with the heat transfer equation for the temperature $T$:

$$Q(x) = -k\nabla^2 T(x) \qquad (26)$$

in order to extract the membrane's thermal profile $T(x)$. In the previous expression $k$ represents the membrane's thermal conductivity.

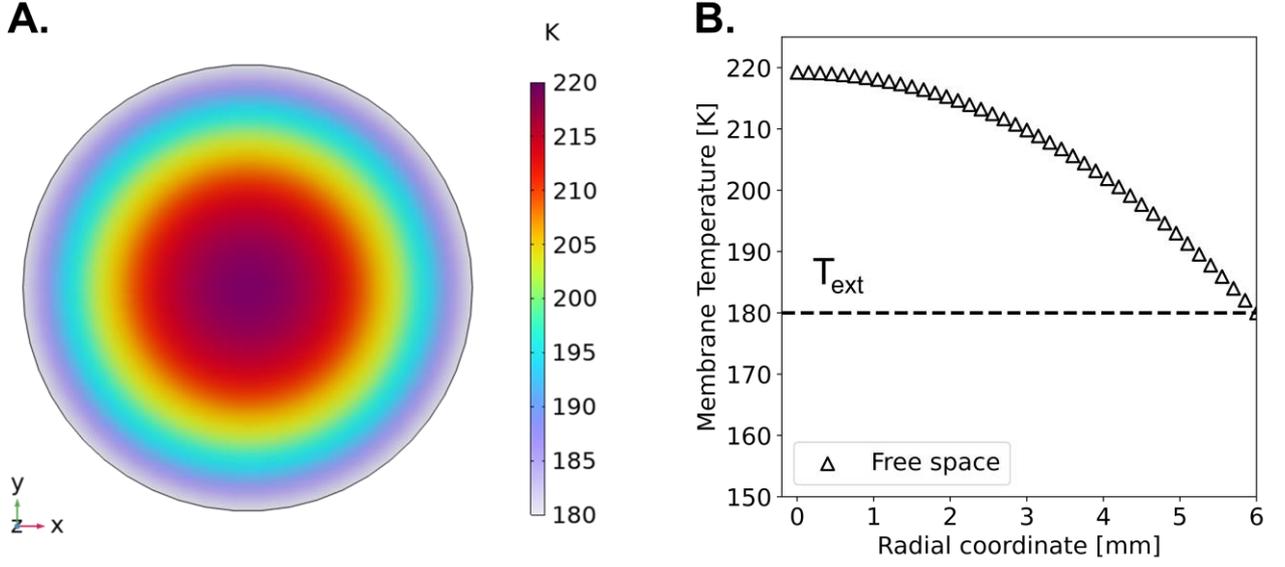

**Extended Data Fig 8: Finite-elements simulation of the membrane's thermal dissipations in free space. A.** Simulated 2D temperature profile of the membrane in free space. **B.** Radial dependence of the membrane's temperature held in free space. The cold-finger temperature has been set at $T_{ext}$ = 180 K.

Firstly, we discuss here the simulated thermal profile of a single silicon nitride membrane held in free space. For simplicity we assumed 2D circular geometry for the membrane. We imposed the boundary conditions in order to have the membrane's edge at the same temperature of the cold finger. The thermal profile along the radial coordinate of the free space membrane will be hence controlled by the balance between the membrane's emissivity ε and the thermal load due to ambient black body radiation at $T_{amb}$ = 300 K. For the simulations we set the silicon nitride emissivity at ε = 0.3 [52] supposing no wavelength dependence across the mid-infrared, where is located the blackbody radiation of the membrane within the employed temperature range (80 - 300 K). Extended Data Fig. 8 illustrates the simulated thermal profile of the membrane in free space, together with a cut along the radial direction.

We highlight that by setting the cold-finger temperature at the temperature at which the phase transition in 1T-TaS$_2$ is observed in free space ($T_{ext}$ = 180 K), we can retrieve a temperature in the middle of the membrane (and hence at the sample position) corresponding to the literature $T_C$. The simulation therefore confirms the assumption that the measured rigid shift of the free space critical temperature (Fig. 1B, C) with respect to the literature one has to be attributed to the high thermal impedance of the Si$_3$N$_4$ membranes between which the sample is embedded, which does not allow them to efficiently re-radiated the ambient blackbody radiation.

Importantly, as shown in Extended Data Fig. 9, the simulations confirm that the incoherent thermal load on the membranes is not significantly influenced by the distance between the cryogenic mirror mounts. This further excludes a trivial scenario in which is the geometrical variation of the cavity mounts which screens the ambient radiation and subsequently changes the membrane's temperature.

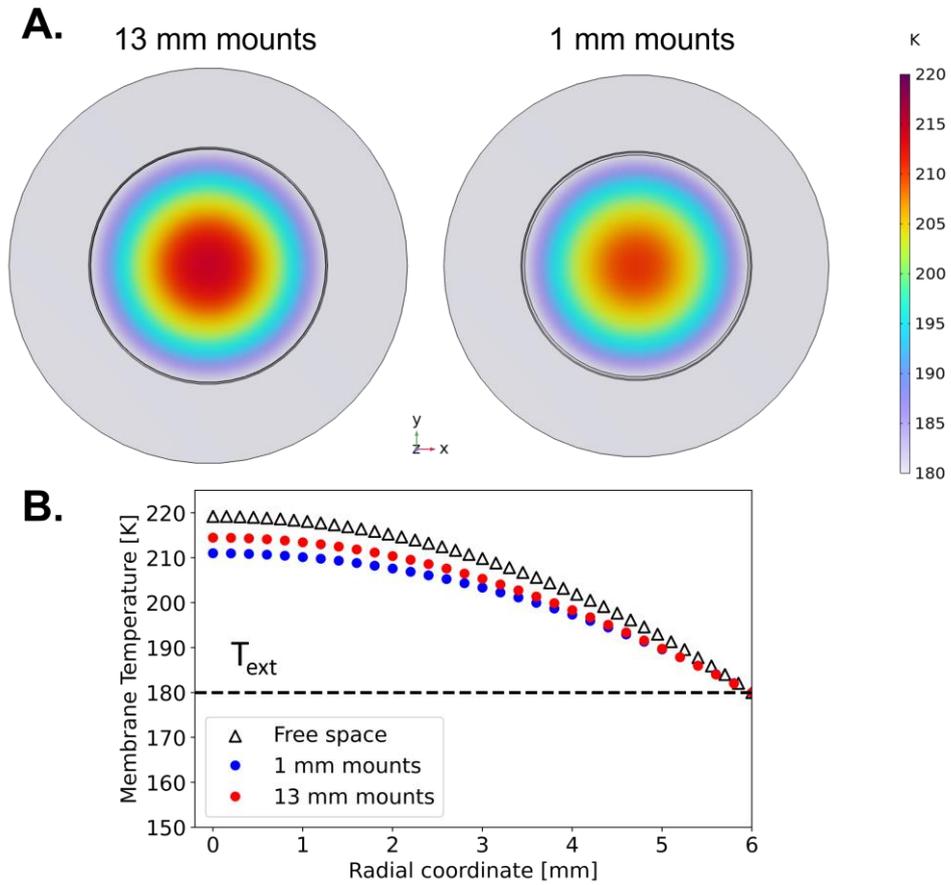

**Extended Data Fig 9: Finite-elements simulation of membrane's temperature as a function of the mirror mounts position. A.** 3D thermal profile of the membrane for two representative distances between the mirror mounts (13 mm and 1.0 mm). **B.** Cut of membrane's thermal profile along the radial direction for the two mounts distances presented in A. The cold-finger temperature has been set at $T_{ext}$ = 180 K, as well as the mounts temperature.

# ACKNOWLEDGMENTS

The authors are indebted to Cesare Tozzo, Paul van Loosdrecht, Hamoon Hedayat, and Omar Abdul-Aziz for insightful discussion. The authors acknowledge Petra Sutar (Jožef Stefan Institute, Ljubljana) for the growth of 1T-TaS$_2$ crystals. The authors acknowledge insightful discussions with Jerome Faist. This work was supported by the European Research Council through the project INCEPT (ERC-2015-STG, Grant No. 677488). F.F. acknowledges financial support from the European Union's H2020 Marie Skłodowska-Curie actions (Grant Agreement No. 799408). D.M. and Petra Sutar acknowledge funding from ARRS grant number P-0040. M.E. acknowledges funding by the Deutsche Forschungsgemeinschaft (DFG, German Research Foundation), Project ID 429529648 – TRR 306 QuCoLiMa ("Quantum Cooperativity of Light and Matter").

# AUTHOR CONTRIBUTIONS

G.J, S.Y.M., and A.M. performed the experiments with support from F.G. and E.M.R.. G.J. analyzed the data with support from S.Y.M.. M.E., D.F., and F.F. conceived and developed the theoretical models with support of D.M. and P.P.. G.J. performed the finite-element simulations of incoherent thermal heating.. R.S. developed the micrometric thermocouple junction for the temperature measurements within the cavity.. S.D.Z. fabricated the silicon nitride membranes and the semi-reflecting cavity mirrors.. D.M. provided the 1T-TaS$_2$ samples.. S.W. provided the THz emitters and guided the THz set-up.. G.J., D.F., A.M., M.E., and F.F. wrote the manuscript with contributions from all the other authors.. D.F. conceived and managed the project.

# DATA AVAILABILITY STATEMENT

Raw hysteretic curves as a function of the cavity frequency (Figure 4A), as well the raw single THz scans of Figs. 4C and 3B are provided in the Supplementary Material. Further datasets collected for this study are available from the corresponding author upon reasonable request.

# COMPETING INTEREST STATEMENT

The authors declare no competing interests.

# SUPPLEMENTARY MATERIAL

1. **Addressing specific questions arising**
   **"Is the observed effect related to incoherent radiation heating or can be associated to the cavity electrodynamics?"**

   a. Does the presence of the sample affect the difference between the temperature of the cold finger and the temperature in the middle of the cavity?
   b. Does the temperature of the cavity mirrors affect the sample temperature?
   c. Does the alignment of the cavity modify the sample temperature?
   d. Does the external radiation influence the sample temperature?
   e. Does the thermal load of the THz radiation affect the observed transition temperature?

2. **Additional datasets**

# 1. ADDRESSING SPECIFIC QUESTIONS ARISING

The main experimental evidence reported in the manuscript is that the effective critical temperature of the metal-to-insulator transition in 1T-TaS$_2$ can be modified by tens of kelvin by placing the sample within a THz cavity. This is linked to a cavity control of the sample temperature, as demonstrated in the main manuscript. We have proved that the observed temperature renormalization depends both on the cavity length and the cavity alignment.

The novelty of the experiment, which represents one of the few attempts in the literature to study how the properties of solid-state complex systems can be modified through the cavity confinement, calls for a thorough characterization of the effect with the aim of ruling out possible experimental artifacts and trivial scenarios. In particular, the most straightforward explanation could be that the huge shift of the effective critical temperature is due to an incoherent radiation heating of the sample, placed within the cavity and therefore in scarce thermal contact with the cold finger.

In the following, we will detail that this is not the case. By jointly discussing results of the finite elements simulations and the temperature measurements, we will give proves pointing towards cavity electrodynamics as the dominant effect explain our experimental observations.

To facilitate the discussion, we will address specific questions that may arise and argue how the complementary tests that we carried out hint towards a cavity-mediated scenario.

### a. Does the presence of the sample affect the difference between the temperature of the cold finger and the temperature in the middle of the cavity?

In a trivial scenario in which the sample experiences an incoherent radiation heating, also the membranes are expected to behave similarly in response to the modification of the cavity geometry. In order to rule out this possibility, we measured the temperature of the sample and the temperature of just the membranes (i.e., without placing the sample between them) by means of the micrometric thermocouple for different cavity fundamental modes.

Fig. S1 presents the results of the temperature measurements performed within the cavity when the thermocouple is sealed on the sample (Fig. S1(A)) and when it is held just between the two membranes (Fig. S1(B)). The temperature measurements, for each cavity setting, are plotted as the difference between the temperature measured on the sample/membrane ($T_{int}$/$T_{membrane}$) and the one recorded on the cold finger ($T_{ext}$) as function of $T_{int}$ or $T_{membrane}$.

We highlight two distinctive trends, characteristic of the presence of the sample:

- The absolute temperature renormalization passing from the lower to higher cavity frequency is significantly higher on the sample with respect to the membrane. For the lowest cold-finger temperature ($T_{ext}$ = 80 K) we indeed measured a renormalization of the sample's temperature of ~ 27 K moving the cavity fundamental mode from 11.5 GHz to 42.8 GHz. Conversely, between the two cavity configurations we measured only a ~ 9 K renormalization of the membrane's temperature.

- The temperature renormalization induced by the cavity in the sample is non-monotonic with respect to the free-space case. Indeed, for high frequency cavities the sample's temperature within the cavity is lower with respect to the sample's temperature measured in free space. This is qualitatively consistent with the trend of the effective critical temperature as function of the cavity fundamental mode revealed by THz spectroscopy (Fig. 4A). Conversely, this anomalous non-monotonic behaviour is not observed on the membranes, where the membrane's temperature measured within the cavity is lower than the free space case for all the cavity frequencies studies. In particular, we observe that the membrane's temperature measured for the larger cavity length (11.5 GHz) approaches the trend measured in free space.

In Fig. S1(C) we present the differential temperature $T_{int} - T_{ext}$ for a fixed cold-finger temperature ($T_{ext}$ = 150 K) as a function of the cavity fundamental frequency. As expected from full temperature scans presented in Fig. S1(A, B) we revealed a renormalization of the sample's effective temperature of ~18 K by sweeping the cavity mode from 11.5 GHz to 42.8 GHz. On the other hand, for the same frequency range, we detected a significantly smaller change in the membrane temperature (~ 5 K).

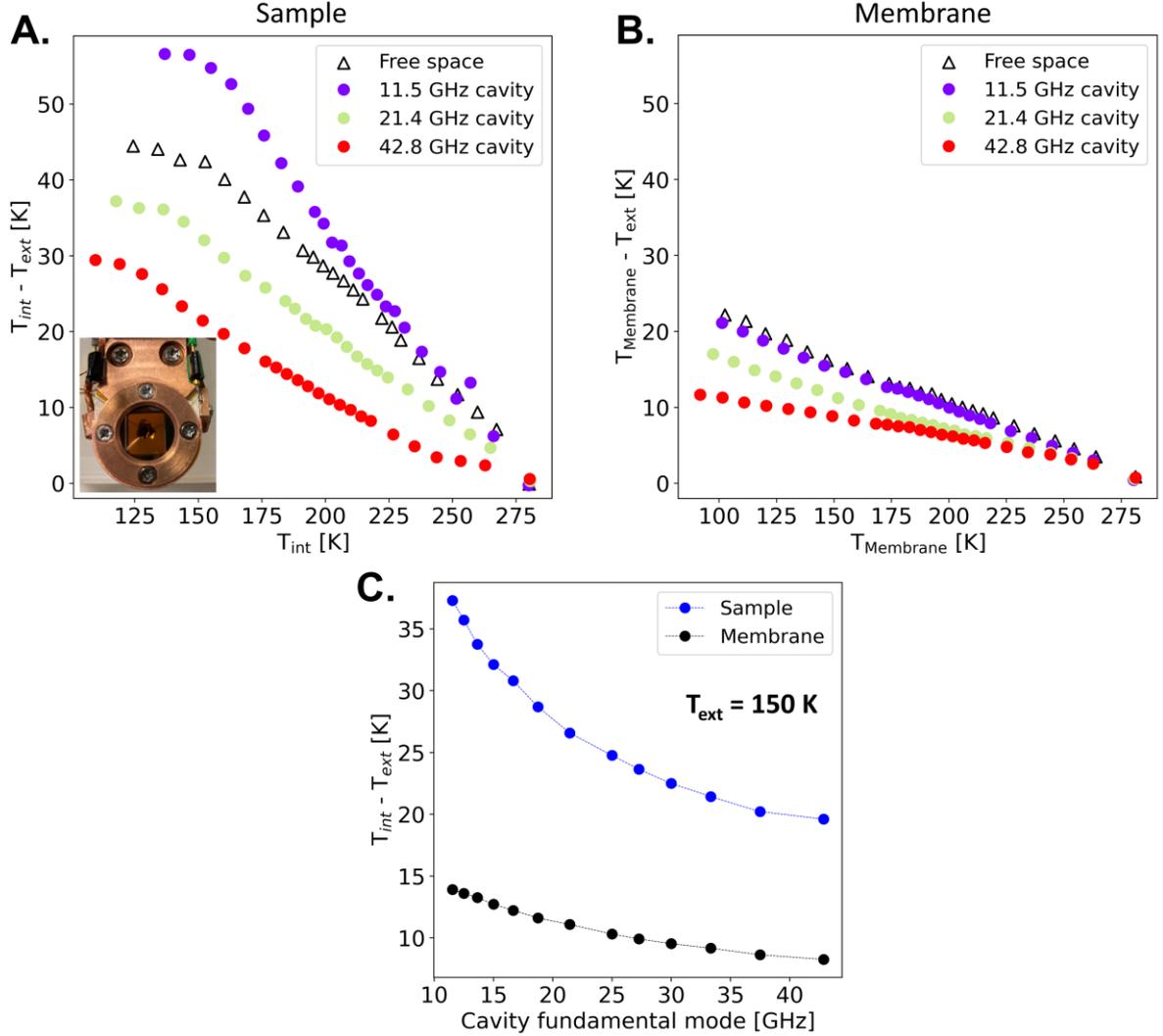

**Fig S1: Temperature measurements within the cavity**. **A.** Difference between the temperature measured on the sample ($T_{int}$) and on the cold finger ($T_{ext}$) as a function of the sample temperature. Temperatures have been measured upon heating the sample from the C-CDW phase. **B.** Difference between the membrane's temperature and the cold-finger temperature when the thermocouple is held just between the two membranes. **C.** Difference between the temperature measured on the thermocouple and on the cold finger as a function of the cavity frequency for a fixed cold-finger temperature ($T_{ext}$ = 150 K). In blue the measurements performed with the thermocouple put on the sample, while in black with the thermocouple held between the membranes.

A further confirmation that the reported effect hints towards a cavity-mediated scenario comes from the direct measurement of the sample's temperature both in heating and cooling conditions. Should the temperature renormalization depend on an incoherent heating, no differences would be expected in the two scanning conditions. This is in fact what we observed on the membranes (Fig. S2(A)), where the difference between $T_{membrane}$ and $T_{ext}$ is identical when heating up or cooling down within the cavity. However, the sample's temperature shows a different trend for the heating/cooling directions (Fig. S2(B)). In particular, while we do not observe any systematic discontinuity when heating up the sample, a kink at the nominal critical temperature ~ 215 K is always present when cooling down the sample at different cavity lengths. In the temperature range ~ 160-215 K we also observe a constant temperature difference between the sample and the cold finger. The thermodynamical meaning of this effect is not clear, but the observation of this hysteretic behaviour, which is not present on the bare membranes, proves that all the effects discussed are peculiar of the sample.

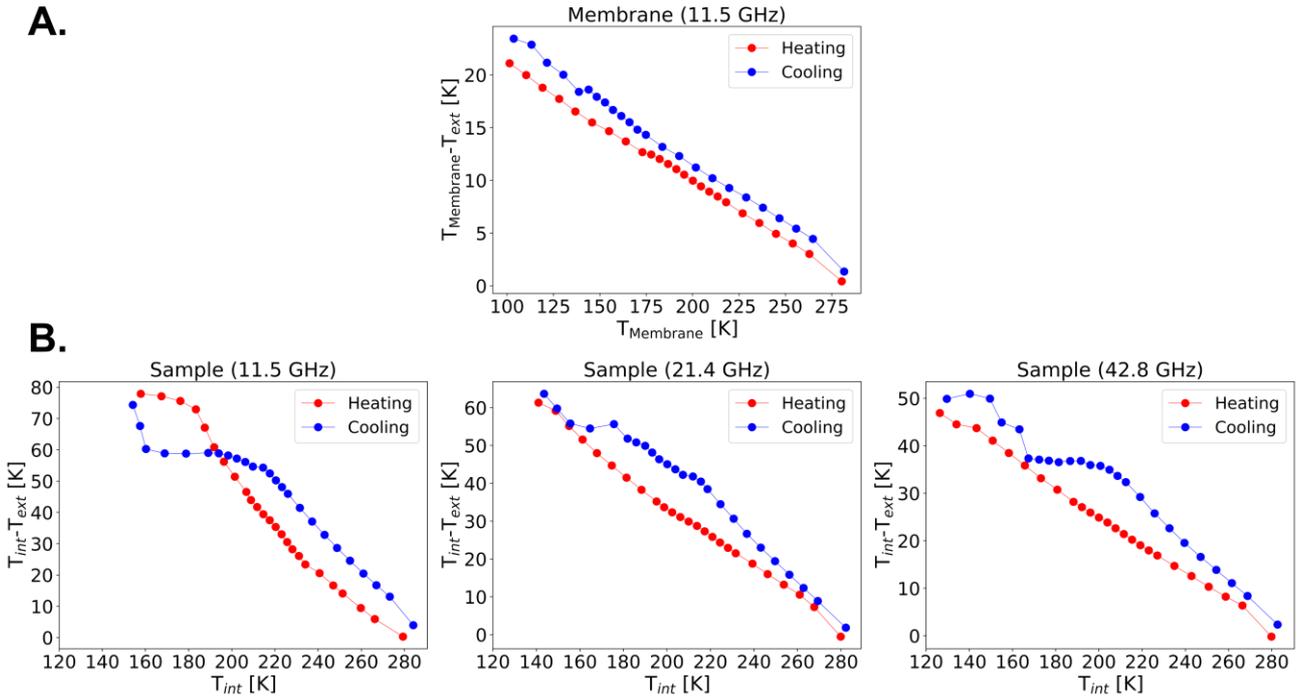

**Fig S2: Temperature measurements upon heating and cooling. A.** Difference between the membrane and the cold-finger temperature within a 11.5 GHz cavity upon heating (red curve) and cooling (blue curve). **B.** Same as (A) but for the sample's temperature in three different cavity configurations.

## b. Does the temperature of the cavity mirrors affect the sample temperature?

A possible source of incoherent thermal load are the cavity mirrors, whose presence might affect the temperature of the sample. In order to exclude this incoherent heating scenario, we carried out a characterization of the phase transition shift with cavity frequency as a function of the mirrors temperature. We made this test by comparing the cryo-cooled (T = 95 K[1]) mirrors configuration with the 290 K mirrors one. The relation between the mirrors and the sample's temperature in the two cases is presented in Fig. S3.

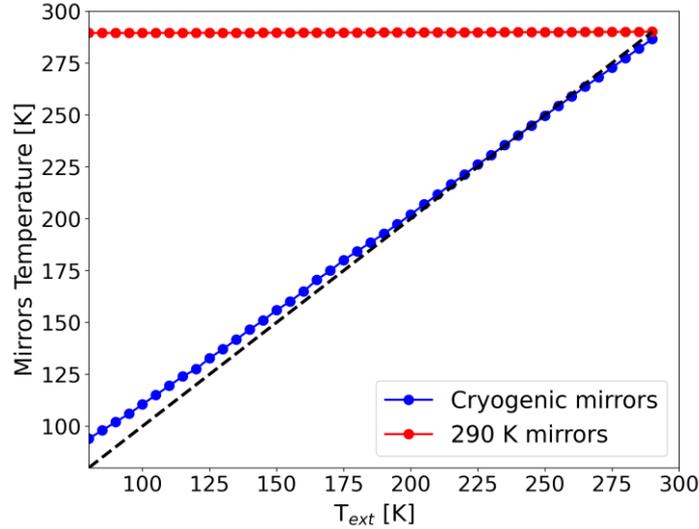

**Fig S3:** Relation between the sample temperature and the mirrors temperature in the cryo-cooled mirrors configuration and in the 290 K mirrors configuration. The black dashed line marks the diagonal.

Fig. S4 shows the temperature dependent THz transmission upon heating 1T-TaS$_2$ embedded in the middle of cavities with three different frequencies ($\omega_c$ = 11.5, 21.2, 48.8 GHz). In Fig. S4(A) the temperature scans were performed with the cavity mirrors at 290 K, while in Fig. S4(B) they were performed in the configuration with cryo-cooled mirrors. The dependence of the effective critical temperature ($T_c^{eff}$) on the cavity frequency is presented in Fig S5 for the two configurations.

It is evident that, despite a rigid temperature shift independent on the cavity frequency (~ 35 K), the effective critical temperature of the phase transition is pushed up with the 290 K mirrors upon increasing the cavity frequency, displaying a trend which is analogous to the one measured with the cryo-cooled mirrors. Crucially, this evidence further hints that the observed shift of $T_c^{eff}$ is an effect due to the cavity confinement. In a trivial incoherent heating scenario, in the 290 K mirrors configuration we would have expected to increase the incoherent thermal load on the sample upon closing the cavity and hence push down the apparent critical temperature of the phase transition.

---

[1] This is the lowest reachable temperature of the mirrors for the present experiment, measured when the cold finger is at $T_{ext}$ = 80 K.

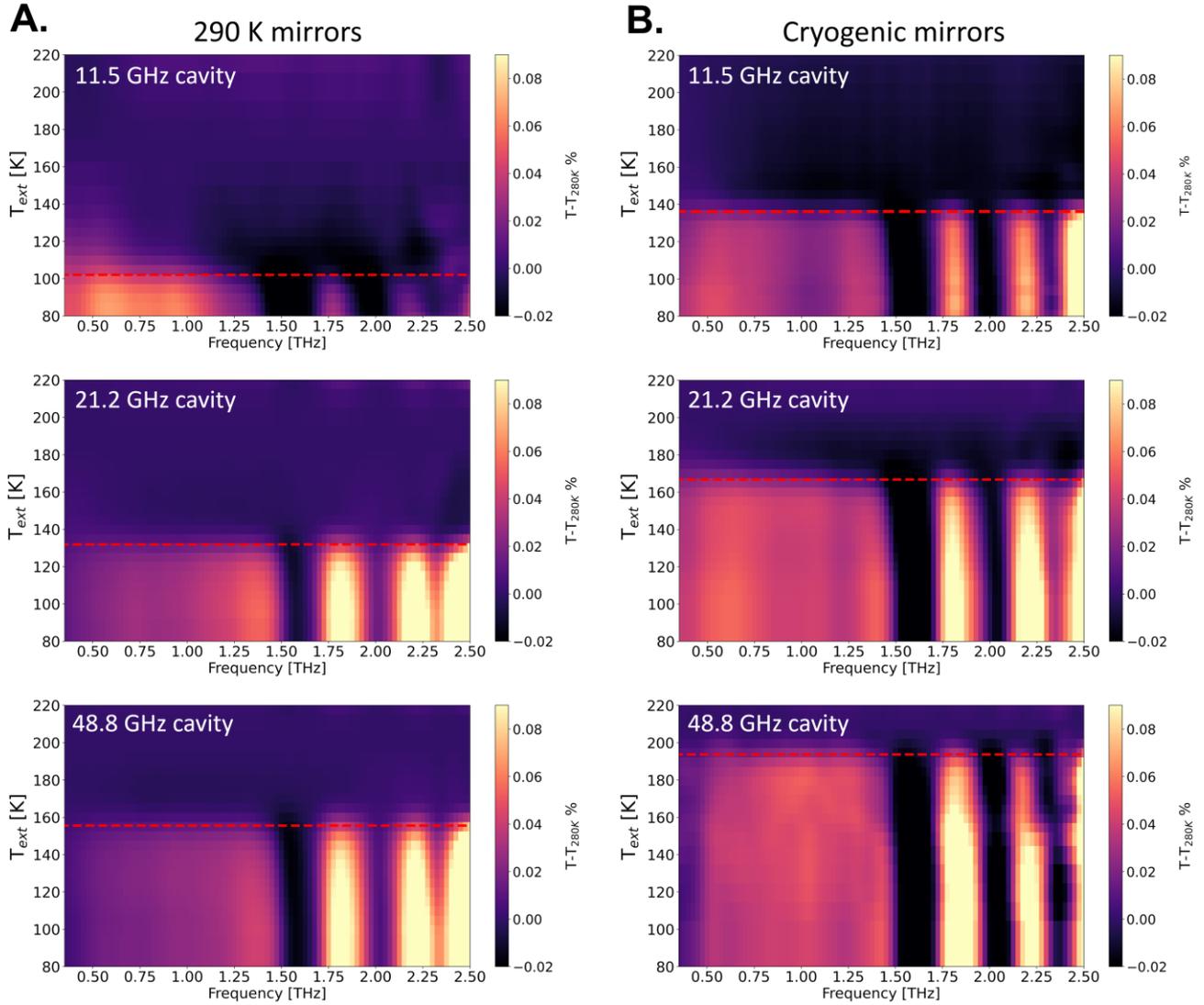

**Fig S4: Dependence of the effective phase transition temperature on the temperature of the cavity mirrors revealed by THz spectroscopy**. **A.** Heating temperature scans for three representative cavity frequencies ($\omega_c$ = 11.5, 21.2, 48.8 GHz) in the 290 K mirrors configuration. **B.** Corresponding heating temperature scans for the cryogenic mirrors configuration.

In order to clarify this point, we simulated with COMSOL the membrane's thermal profile embedded in three cavities having different lengths. The employed 3D geometry can be found in Fig. S6(A) and Fig. S7(A) for cryogenic and ambient temperature mirrors, respectively.
The radial profiles plotted in Fig. S6(C) and Fig. S7(C) show that, upon closing the cavity and thus increasing the cavity frequency, the shielding of ambient radiation is efficient only with the cryogenic mirrors, whereas in the 300 K mirror case the incoherent radiation from the mirrors surfaces dominates the thermal load on the membrane. We stress that this trend is opposite to the one measured by THz spectroscopy (Fig. S4, S5).

Together with THz transmission, we characterized the effect of the mirrors temperature on the phase transition by tracking the sample's temperature as a function of the cavity length and of the cavity alignment. Fig. S8 shows a comparison of the differential temperature between the cold finger and the sample when the latter is held between the cryogenic cavity (Fig. S8(A)) and between the cavity with 290 K mirrors (Fig. S8(B)). The measured trend is qualitatively consistent with the effective critical temperature trend measured by THz spectroscopy (Fig. S5). Indeed, despite a cavity-independent shift of the sample's temperature due to the incoherent thermal load, in the 290 K mirrors case the sample's temperature is pushed down with a similar trend of the cryogenic mirrors case upon increasing the cavity frequency.

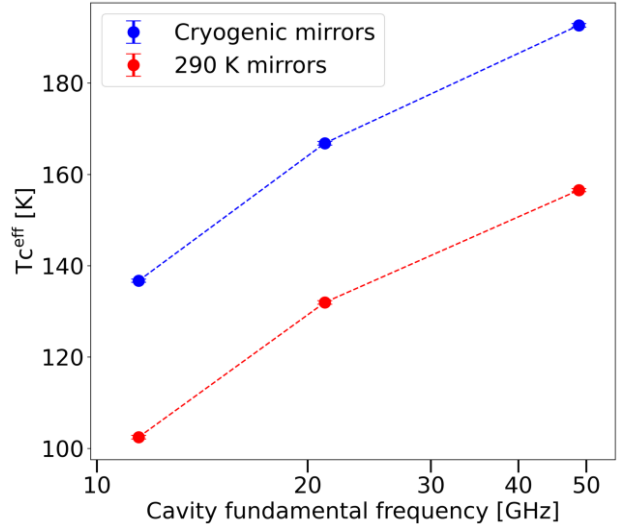

**Fig S5: Dependence of the heating critical temperature on the cavity fundamental frequency for the 290 K and cryogenic mirrors configurations.** In blue (red) the effective critical temperature measured upon heating the sample from the dielectric phase for the cryogenic mirrors (290 K mirrors) configuration.

For the lowest cold-finger temperature ($T_{ext}$ = 80 K) and in the 290 K mirrors case we measured a renormalization of the sample's temperature of ~ 31 K by sweeping the cavity mode from 11.5 GHz to 42.8 GHz, which is similar to the ~ 27 K measured within the cryogenic cavity. A similar trend between the 290 K and cryo-cooled cases is measured by fixing the cold finger at 150 K and tracking the differential temperature $T_{int} - T_{ext}$ as a function of the cavity fundamental mode (Fig. S8(C)). As highlighted in the comparative plots of Fig. S8(C), the renormalization of the sample's temperature measured either within the cryogenic cavity and with the 290 K mirrors is not consistent with the trend measured on the bare membranes, where a ~ 3 times smaller renormalization is observed moving the cavity mode at 150 K from 11.5 GHz to 42.8 GHz (black points of Fig. S8(C)).

Crucially, we further proved that the cavity-mediated change of sample's temperature is independent on the mirrors temperature by repeating the misalignment test with the 290 K mirrors. In Fig. S9 we present the comparison of the differential temperature $T_{int} - T_{ext}$ as function of the cavity alignment in the cryo-cooled and in the 290 K mirrors case. We show that, despite a rigid shift due to the incoherent thermal load introduced by the 290 K mirrors, the renormalization of the sample's temperature due to the cavity alignment is ~ 3.5 times larger than the one measured on the bare membranes for both the mirrors temperature configurations. This evidence further excludes that the observed effect is related to the mirror's incoherent heating and hints to a selective effect of the cavity electrodynamics on the sample's temperature.

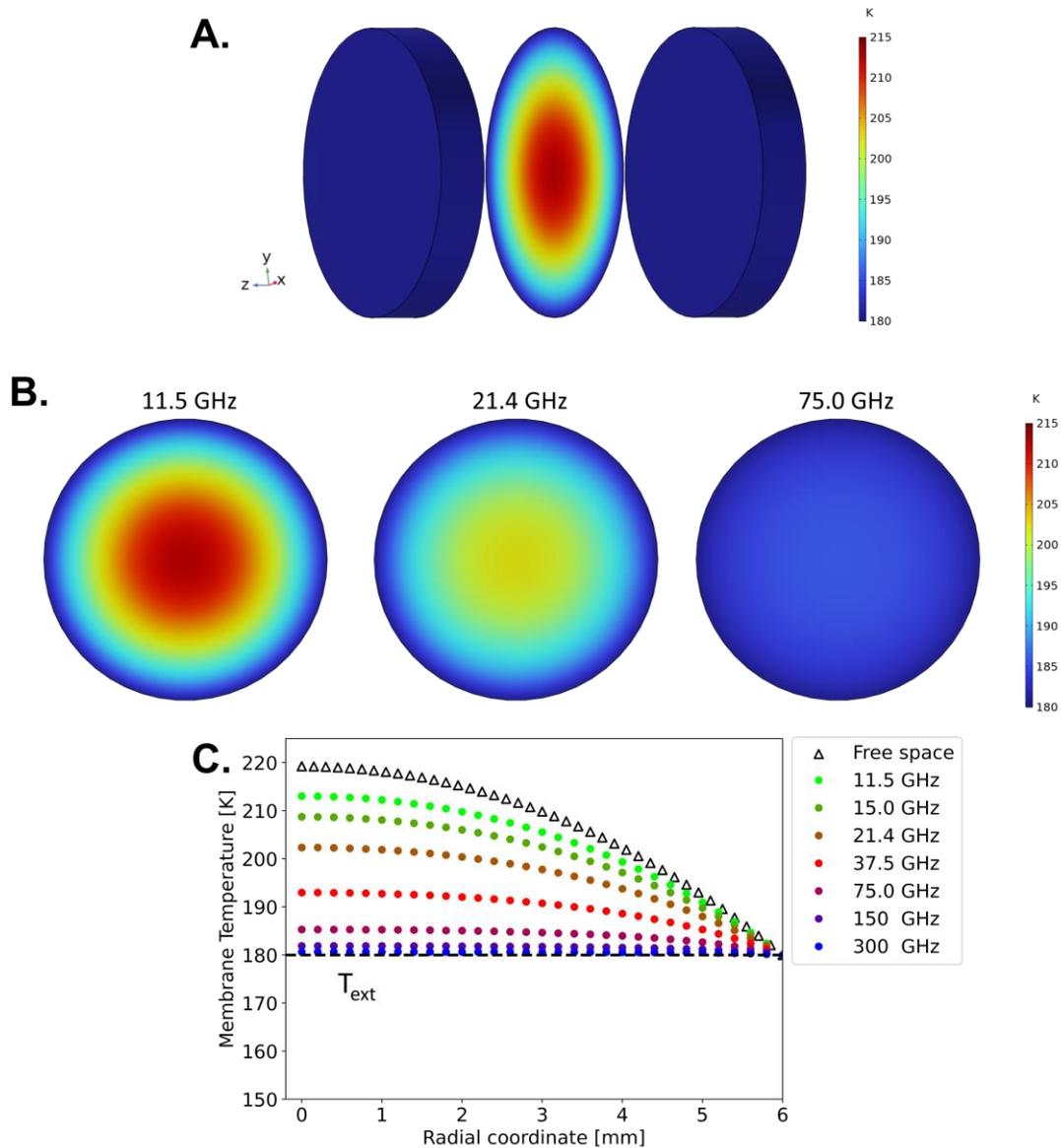

**Fig S6: Finite elements simulation of membrane's temperature as a function of the cavity fundamental frequency in the cryogenic mirrors configuration. A.** 3D thermal view of the cryogenic cavity employed for the simulations. **B.** 2D thermal profile of the membrane within the cryogenic cavity for three representative cavity frequencies (11.5, 21.4, 75.0 GHz) employed in the experiment. **C.** Radial profile of the membranes held within the cryogenic cavity for different values of the cavity fundamental mode (indicated in legend). The cold-finger temperature ($T_{ext}$) and the mirrors temperature have been set at 180 K.

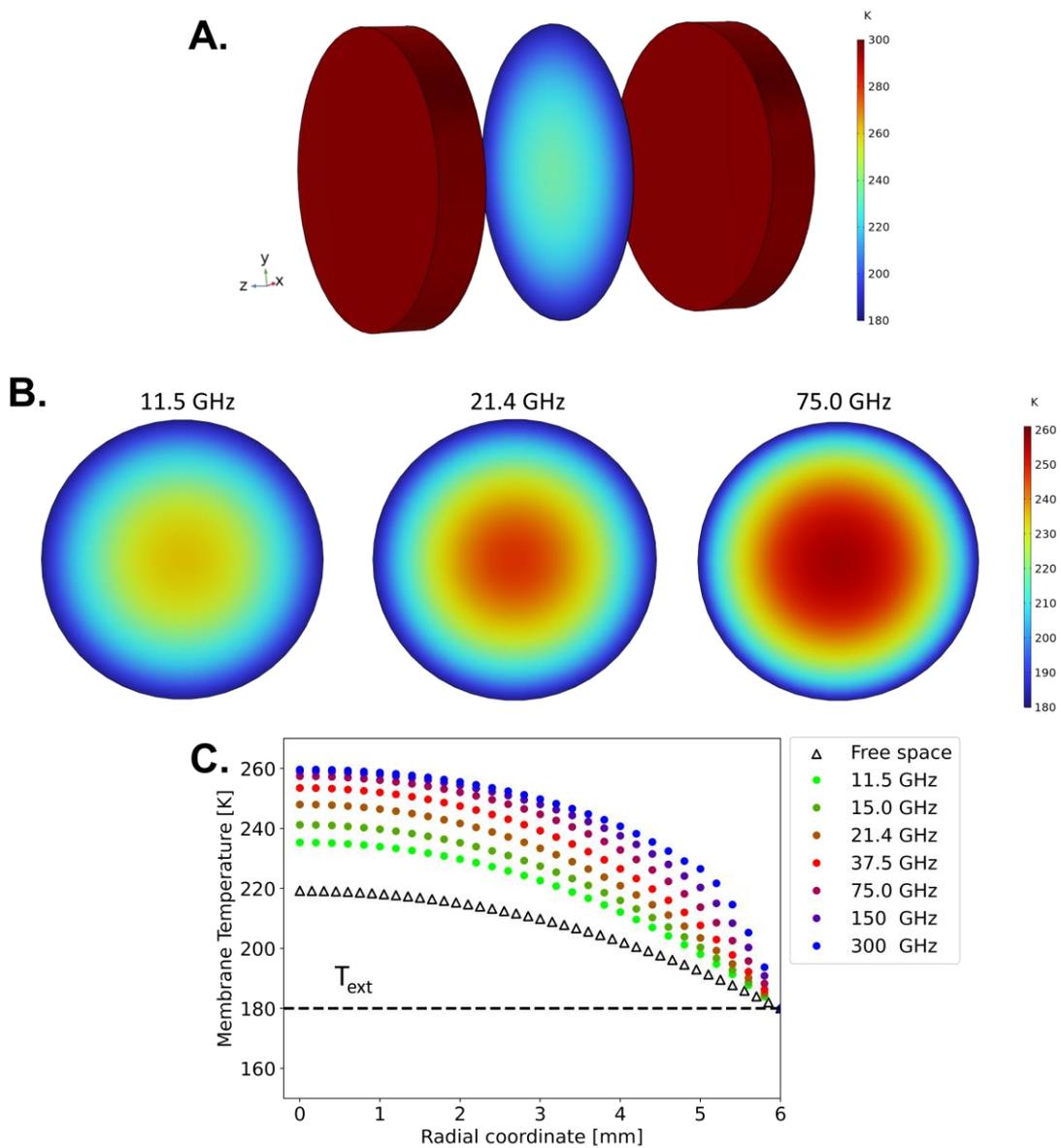

**Fig S7: Finite elements simulation of membrane's temperature as a function of the cavity fundamental frequency in the 300 K mirrors configuration. A.** 3D thermal view of the membrane held between the 300 K mirrors. **B.** 2D thermal profile of the membrane within the 300 K mirrors for three representative cavity frequencies (11.5, 21.4, 75.0 GHz) employed in the experiment. **C.** Radial profile of the membranes in the 300 K mirrors configuration for different values of the cavity fundamental mode (indicated in legend). The cold-finger temperature ($T_{ext}$) has been set to 180 K.

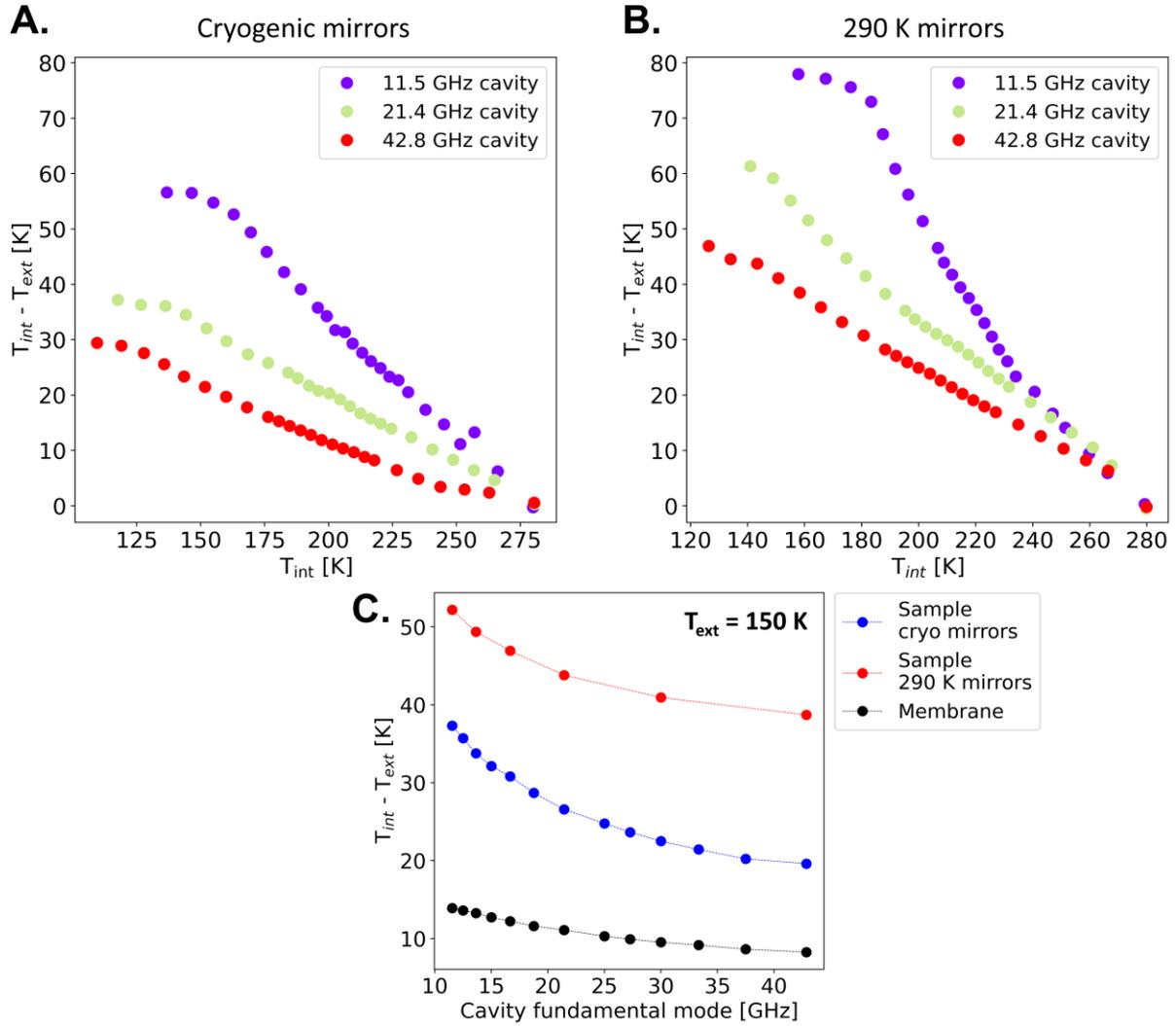

**Fig S8: Temperature measurement within the cavity as a function of the mirrors temperature. A.** Evolution of the difference between the temperature measured on the sample ($T_{int}$) and on the cold finger ($T_{ext}$) as a function of the sample temperature in the cryo-cooled mirrors configuration. **B.** Differential temperature $T_{int} - T_{ext}$ as a function of the sample's temperature measured in the 290 K mirrors configuration. In both cases the temperatures have been measured upon heating the sample from the C-CDW phase. **C.** Differential temperature $T_{int} - T_{ext}$ for a fixed cold-finger temperature ($T_{ext}$ = 150 K) as a function of the cavity fundamental frequency. In blue the measurements performed within the cavity with cryogenic mirrors, while in red with 290 K mirrors. In black, for reference, the differential temperature measured within the membranes in the cryogenic mirrors configuration.

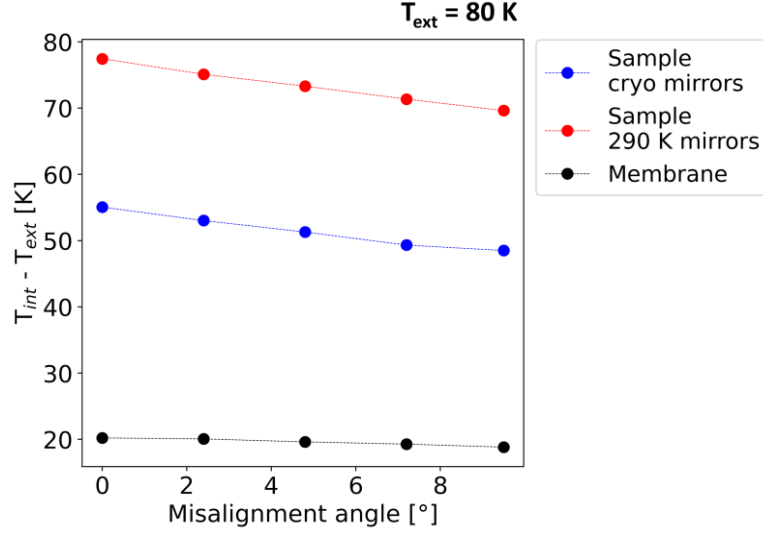

**Fig S9: Temperature measurements within the cavity as a function of the cavity alignment for different mirror temperatures.** Differential temperature $T_{int} - T_{ext}$ as a function of the alignment measured for the sample held within cryogenic mirrors (blue) and 290 K mirrors (red). In black, for comparison, the differential temperature measured on the membranes. For the present measurements the cold finger has been set at $T_{ext} = 80$ K.

### c. Does the alignment of the cavity modify the sample temperature?

A further parameter which was shown to affect the effective critical temperature of the metal-to-insulator transition is the alignment of the cavity, which ultimately sets its quality factor (Fig. 3B of the main manuscript). In order to demonstrate that this is not just a pure geometrical effect, we characterize in the following the response of the sample to the cavity misalignment.

First of all, we proved that the effective change of the phase transition temperature due to the cavity alignment cannot be ascribed to a change of the cavity length. In order to do so, we estimated the change of the cavity frequency at the sample position as a function of the misalignment angle of the mirrors. Fig. S10(A) illustrates the THz time-domain traces in the C-CDW phase for different misalignment angles Θ. In the dashed box we highlight the THz reflection associated to the cavity round trip whose temporal distance from the main transmitted peak sets the cavity length. We note that upon misalignment the peak associated to the cavity round trip reduces its intensity. This is consistent with the decrease of the photon lifetime within the cavity and hence with the reduction of the quality factor of the bare cavity, which has been estimated in Fig. S10(B) as a function of the misalignment angle by approximating the exponential decay with a linear fit.

The dependence of the cavity frequency shift on Θ and the corresponding linear fit are shown in Fig. S10(C). We estimated the change of the fundamental frequency upon misalignment to be $\Delta\omega_c = 0.14$ GHz/deg for the 11.5 GHz cavity. This implies that the 11.5 GHz cavity misaligned at the maximum angle (Θ = 9.5°) has an equivalent fundamental frequency of 12.8 GHz. We therefore set the cavity frequency at 12.8 GHz and compare the effective critical temperature obtained in this configuration with the one of the sample embedded within the 11.5 GHz cavity misaligned at Θ = 9.5°. We measured the shift of the effective critical temperature between the two cavity configurations to be $\Delta_{MIS} = 21$ K. The apparent shift of the phase transition can be hence attributed to the cavity misalignment (Fig. S10(D)), as it cannot be justified just in terms of a change in the cavity length.

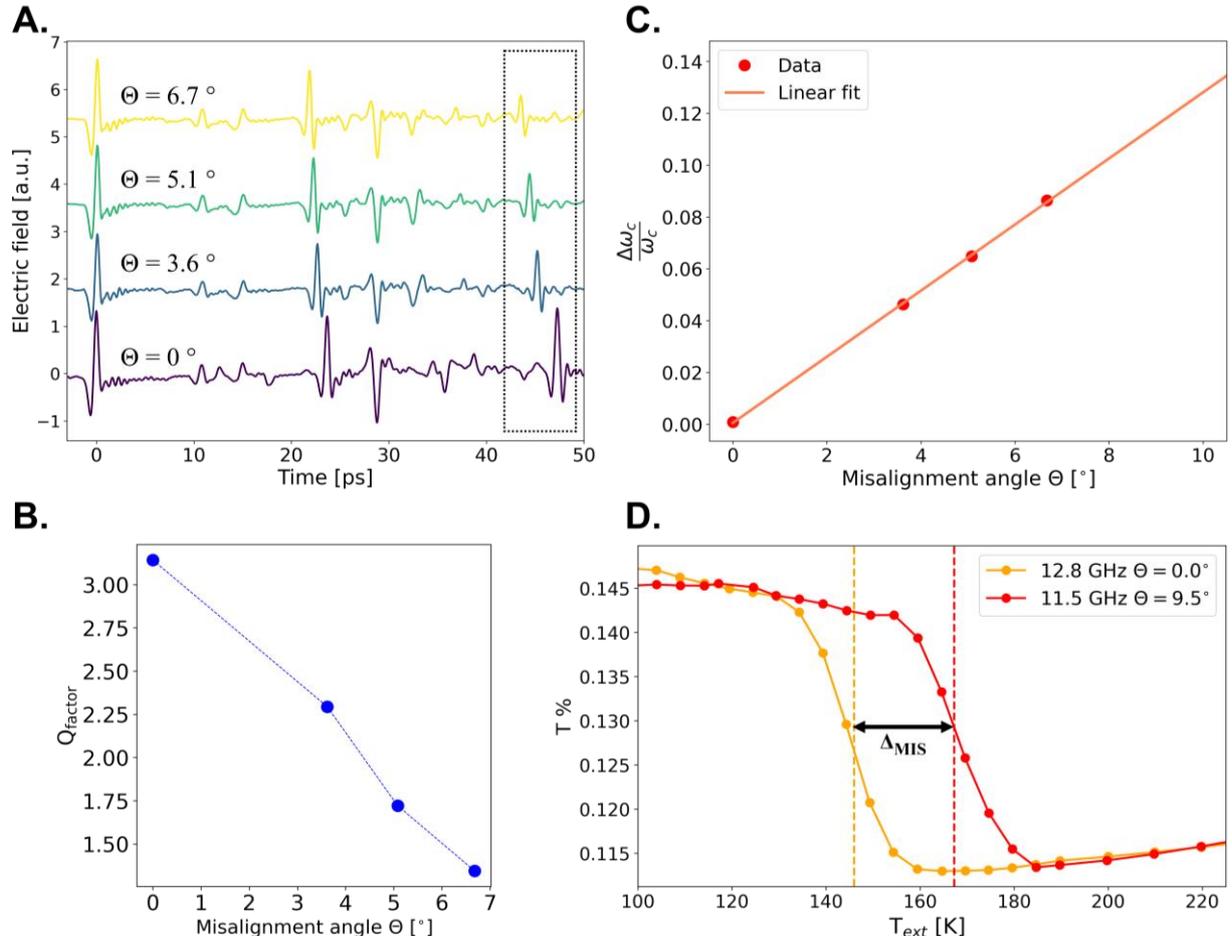

**Fig S10: Variation of the cavity fundamental frequency as a function of the total misalignment angle of the cavity**. **A.** THz time domain fields passing through the sample within the cavity for different misalignment angles Θ. In the dashed box we highlight the THz reflection associated to the cavity round trip. **B.** Estimated cavity quality factor as a function of the misalignment angle. **C.** Relative shift of the cavity frequency as a function of the misalignment angle obtained from the THz fields shown in A and corresponding linear fit. **D.** Comparison between the temperature dependent low frequency transmission (0.2 THz < ω < 1.5 THz) in the 11.5 GHz misaligned cavity (Θ = 9.5°) and in the 12.8 GHz aligned one. The measured temperature shift $\Delta_{MIS}$ = 21 K quantifies the shift of the effective critical temperature $T_c^{eff}$ due to cavity misalignment.

Furthermore, by means of the finite elements simulations, we proved that the renormalization of the effective critical temperature cannot be explained by simply assuming an incoherent thermal heating. In fact, as shown in Fig. S11, no temperature shift in the membrane is expected when the cavity mirrors are in the maximum misaligned configuration considered in the experiment (θ = 9.5°).

Finally, in order to verify how the sample's temperature is affected by the cavity environment, we measured $T_{int}$ as a function of the cavity alignment for a cavity fundamental mode of 11.5 GHz (Fig. S12). By setting the cold-finger temperature to 80 K, we detected a renormalization of 6.5 K in the sample's temperature by switching from the aligned to the maximum misaligned condition. Conversely, we measured only a 1.6 K temperature variation when the thermocouple is held only between the two membranes. We note that the measured renormalization of the sample's temperature is much smaller than the shift of the effective transition temperature measured by THz spectroscopy (Fig. 3 in the main).

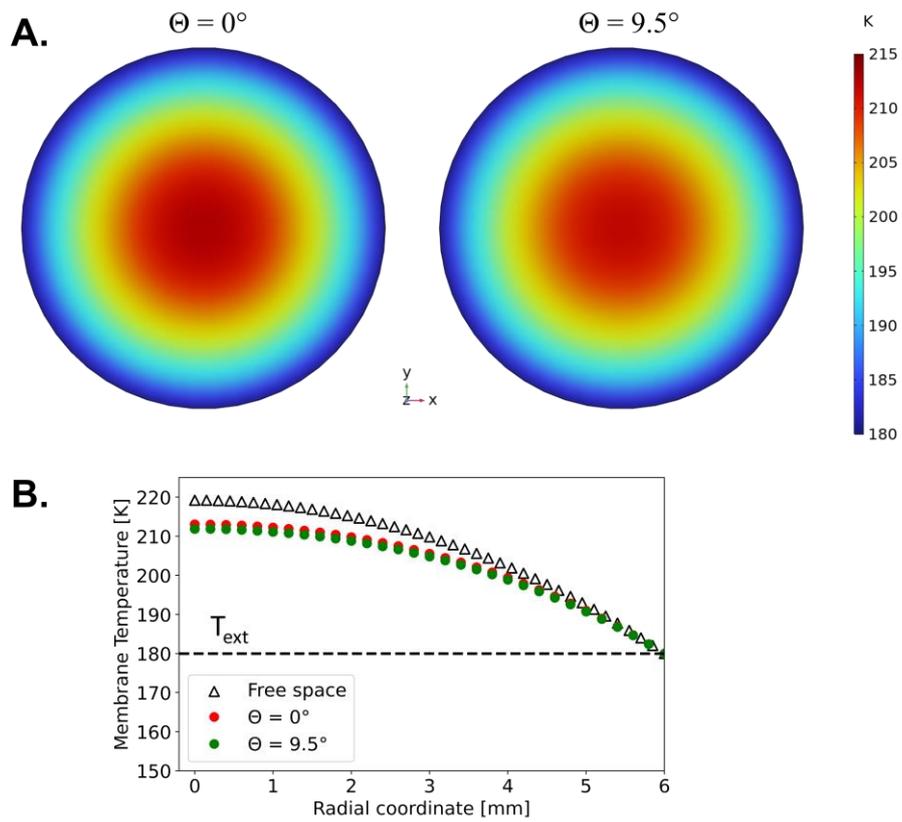

**Fig S11: Finite elements simulation of the membrane's temperature as a function of the cavity alignment. A.** 2D thermal profile of the membrane within the cryogenic cavity in the aligned configuration ($\theta = 0°$) and in the maximum misaligned configuration employed in the experiment ($\theta = 9.5°$). **B.** Membrane's thermal profile along the radial coordinate for the two alignment conditions. The cold-finger temperature ($T_{ext}$) and the mirrors temperature have been set to 180 K.

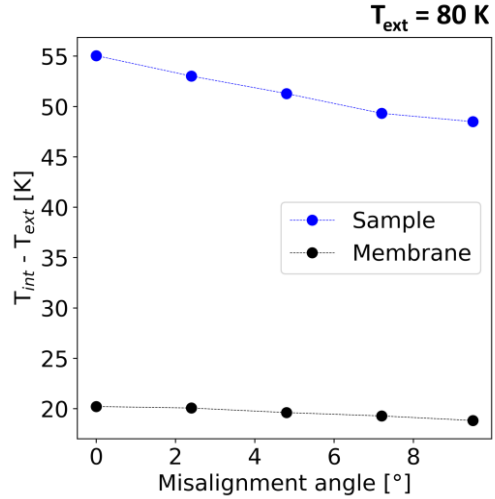

**Fig S12: Temperature measurements within the cavity (11.5 GHz) as a function of the cavity alignment.** In blue the dependence of the difference between the temperature measured on the sample ($T_{int}$) and on the cold finger ($T_{ext}$) as a function of the mirror alignment. In black, for comparison, the same differential temperature $T_{int} - T_{ext}$ measured on the membranes. For the presented measurements we set the cold finger at $T_{ext}$ = 80 K.

### d. Does the external radiation influence the sample temperature?

In order to prove that the shift of the effective critical temperature upon tuning the cavity resonance is not an effect merely due to the geometry of the cavity chamber, we removed the cavity mirrors and tracked the phase transition of the sample in free space at two different positions of the mirrors mounts (Fig. S13). For this characterization we compared the mirrors mounts distance corresponding to a 9 mm (16.7 GHz) cavity with the one corresponding to a 1 mm (150 GHz) cavity. No significant shift of the effective critical temperature (~ 2.0 K) is measured between the two configurations. This proves that the thermal load on the sample is not influenced by the distance from the cryogenic mirrors mounts, and hence that the effective critical temperature shift upon tuning the cavity mode (Fig. 4A of the main manuscript) cannot be trivially ascribed to a geometrical variation of the cavity chamber. Therefore, the critical temperatures measured in free space discussed in the main manuscript (Fig. 1) effectively set the absolute free space reference for all the cavity-dependent studies.

We also demonstrated that the leading effect is not related to a geometrical screening of the black-body ambient radiation, whose amount within the cavity can be geometrically modified by tuning the mirrors distance. To exclude this scenario, we screened with metallic foils the cavity chamber and, by means of the micrometric Cr-Al junction, tracked the differential temperature between the sample and the cold finger as a function of the cavity frequency. Importantly, as highlighted in Fig. S14, a similar temperature trend is detected in both the shielded and non-shielded configurations. This, together with the evidence of Fig. S13, further validates that the reported evidence cannot be simply explained in terms of a geometrical screening of the ambient blackbody radiation and more likely hints at a scenario mediated by cavity electrodynamics.

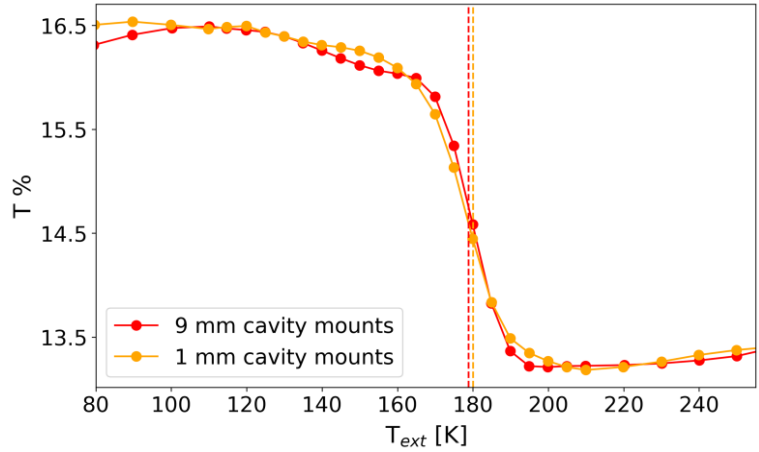

**Fig S13: Dependence of the effective phase transition temperature on the cavity geometry (without mirrors)**. The low frequency transmission (0.2 THz < ω < 1.5 THz) in free space is plotted for two representative distances between the mirrors mounts and the sample. No significant shift in the effective critical temperature is observed (~ 2 K).

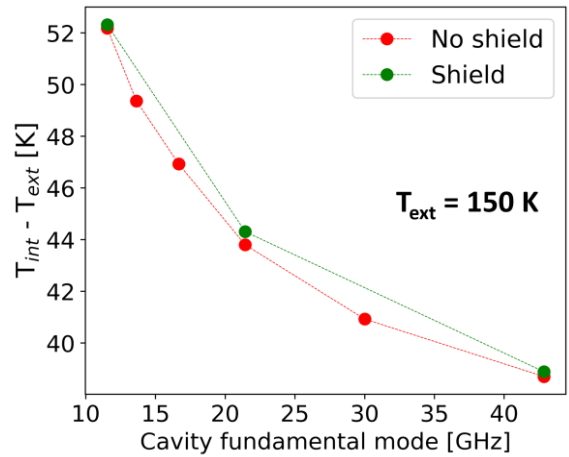

**Fig S14: Effect of the shielding of ambient radiation on the sample's temperature**. Differential temperature between the sample and the cold finger as a function of the cavity fundamental mode. Red (green) points correspond to sample temperatures measured without (with) shielding the cavity environment with alumina foils. The test has been made with the mirrors temperature set at 290 K.

e. **Does the thermal load of the THz radiation affect the observed transition temperature?**

In order to verify that the phase transition within the cavity is not influenced by the thermal load introduced by the THz radiation, we repeated the same temperature scan with different intensities of the THz pulse. This was achieved by varying the bias voltage of the photoconductive antenna. Fig S15 shows the cooling temperature scans for the sample within a cavity of a representative frequency of 36.8 GHz for two different peak strengths of the input THz field (0.1 KV/cm, and 0.03 KV/cm). A negligible shift of the effective critical temperature (< 1.0 K) is measured, confirming that the employed THz pulse only acts as probe and does not introduce a detectable thermal load at the sample position. For this reason, in order to maximize the signal to noise ratio of the detected THz field, all the measurements presented were performed at a THz input peak strength of 0.1 KV/cm.

A further confirmation of the negligible effect of the THz thermal load on the reported evidence is given by directly measuring the temperature of the membrane in the absence and in the presence of the THz field. In Fig. S16, we plot the temperature difference between the cold finger and the centre of the membrane (measured by means of the Cr-Al junction). We measured that even with the maximum THz intensity employed in all the measurements in the manuscript (0.1 kV/cm), the temperature difference $T_{Membrane} - T_{ext}$ changes by less than 1 K.

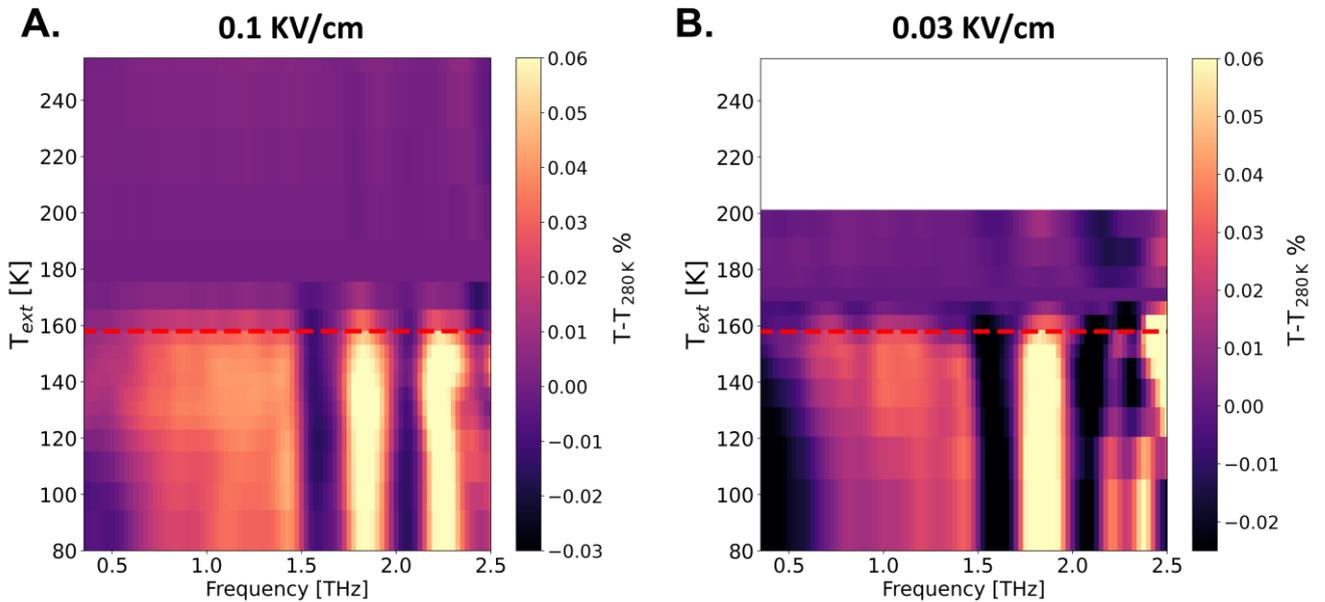

**Fig S15: Dependence of the metal-to-insulator transition on THz probing intensity.** Cooling temperature scans for the sample within a cavity of a representative frequency of 36.8 GHz for two different intensities of the THz probing field: 0.1 KV/cm (**A**) and 0.03 KV/cm (**B**). A negligible shift of the effective critical temperature (< 1.0 K) between the two THz intensities is detected.

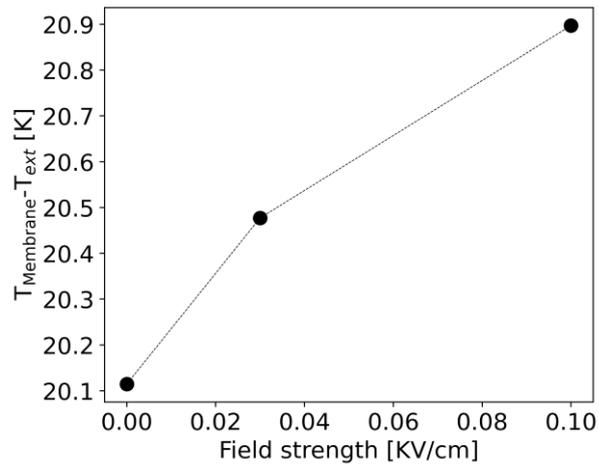

**Fig S16: Dependence of the membrane temperature on the THz intensity**. Measured difference between the temperature of the membrane and the temperature of the cold finger in the absence (0 kV/cm) and in the presence of THz radiation with two different intensities (0.03 kV/cm and 0.1 kV/cm).

# 2. ADDITIONAL DATASETS

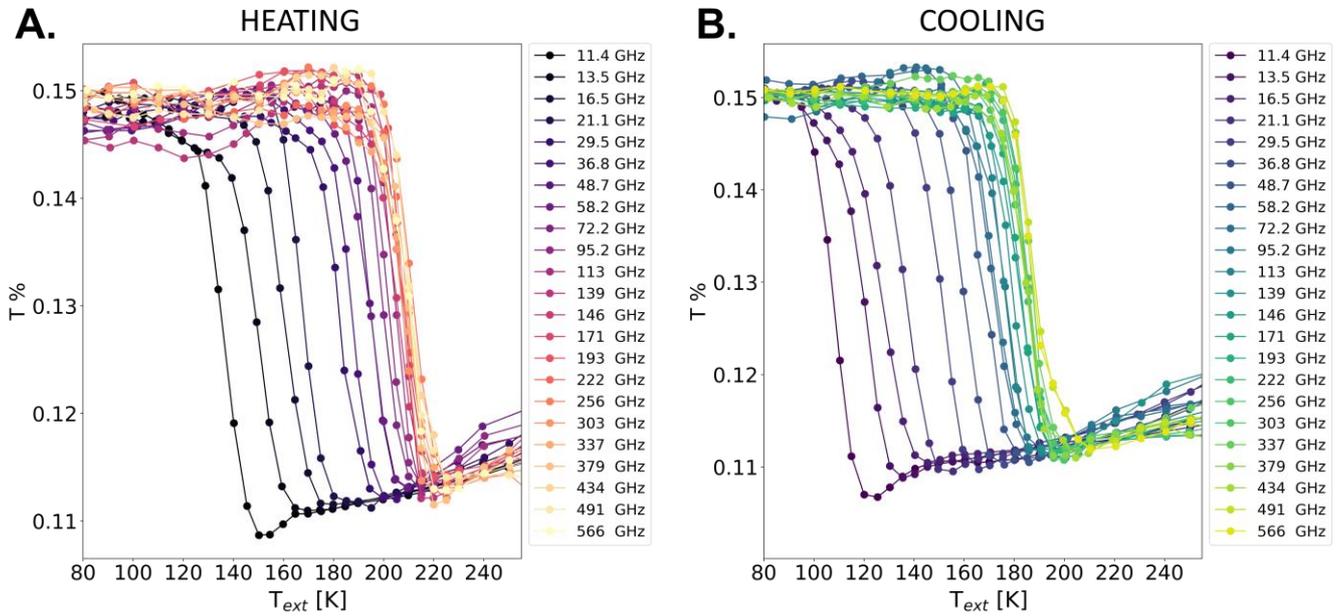

**Fig S17: Raw temperature hysteretic curves as a function of the cavity frequency. A.** Low frequency THz transmission (0.2 < ω < 1.5 THz) for all the cavity frequencies of Fig. 4A (indicated in legend) measured upon heating the sample. **B.** Corresponding sweep curves for the cooling temperature scans.

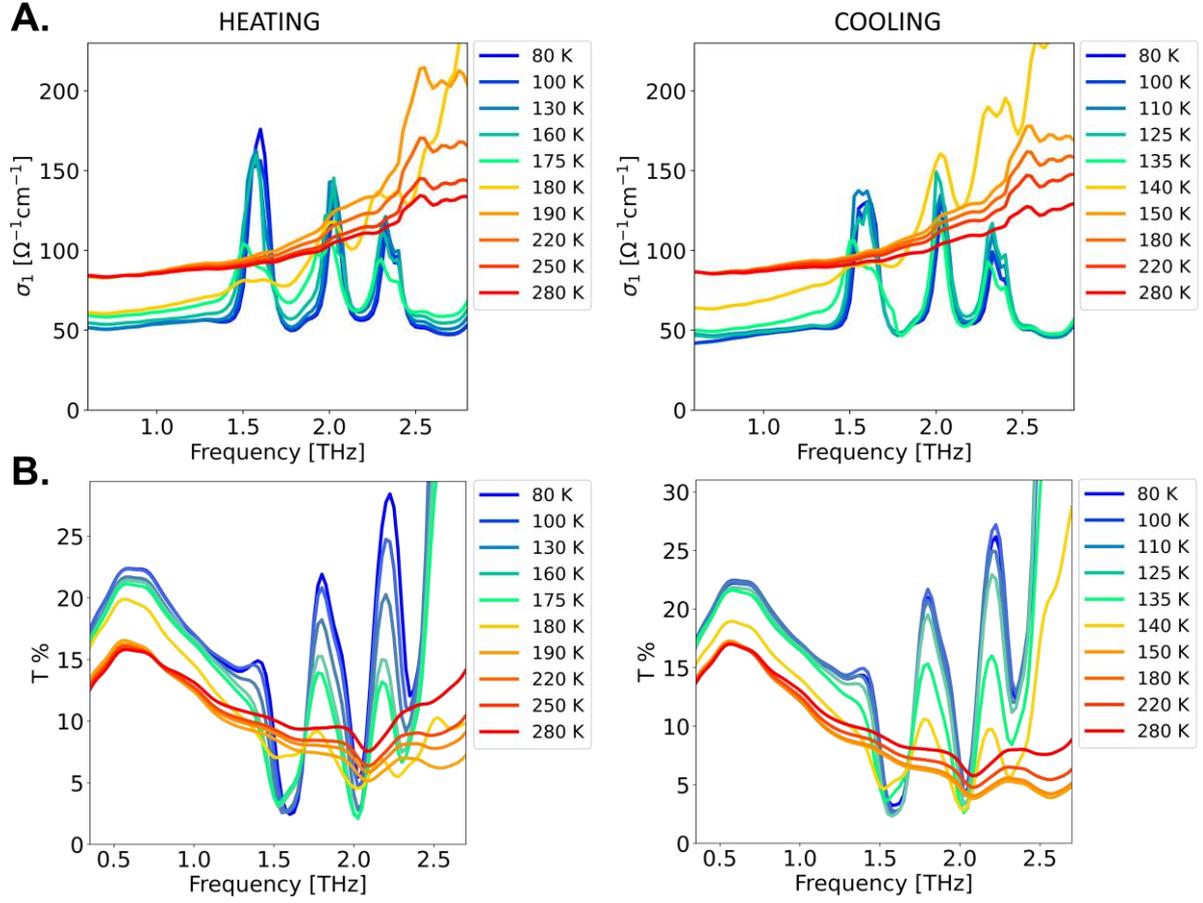

**Fig S18: Optical conductivity and THz transmission in 1T-TaS$_2$ across the metal-to-insulator phase transition. A**. Real part of the optical conductivity ($\sigma_1(\omega)$) measured in free space upon heating (left panel) and cooling (right panel) the sample from the insulating and metallic phase, respectively. **B**. THz transmission (T) measured in free space at different cold-finger temperatures $T_{ext}$ upon heating (left panel) and cooling (right panel) the sample.

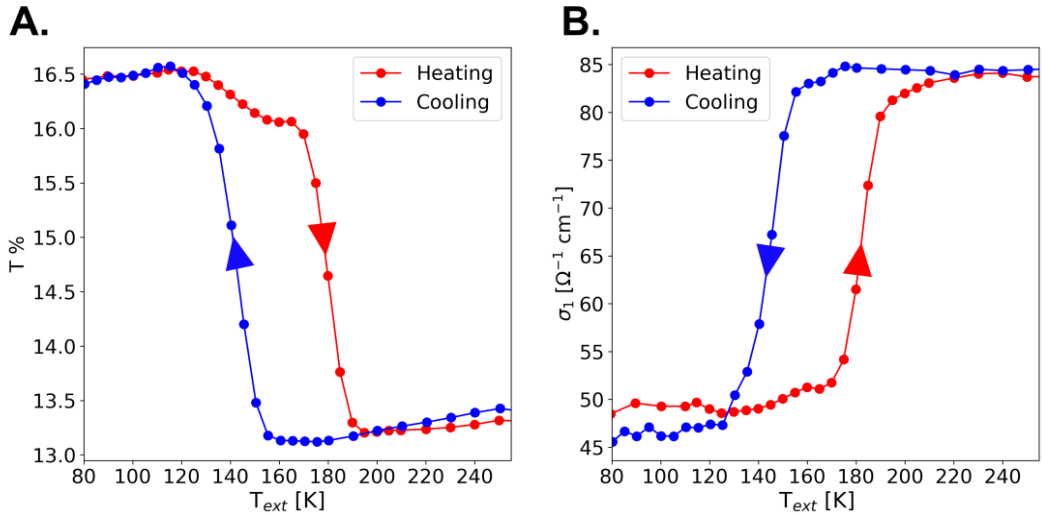

**Fig S19: Hysteretic behaviour in THz transmission and optical conductivity. A**. Low-frequency transmission ($0.2 < \omega < 1.5$ THz) upon heating and cooling the sample. **B**. Real part of the optical conductivity integrated in the same frequency range upon heating and cooling. The hysteretic behaviour is the same for both observables.

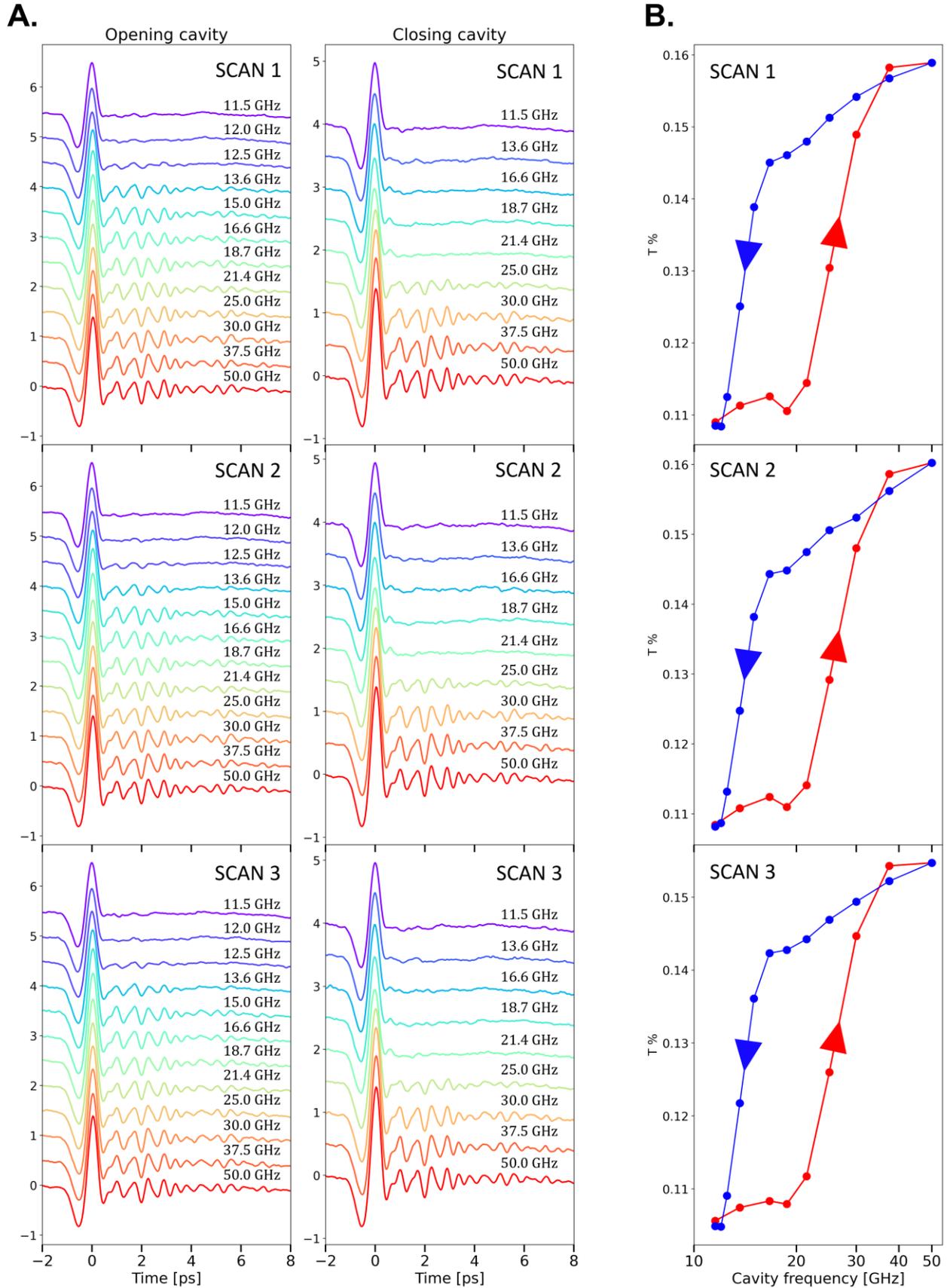

**Fig S20: Single scans as function of the cavity length. A**. Single THz fields measured upon opening (left) and closing (right) the cavity at a fixed temperature (150 K). The 3 consecutive scans employed for the average (insets of Fig. 3B) are presented. **B**. Corresponding low frequency transmission ($0.2 < \omega < 1.5$ THz) as a function of the cavity frequency for three consecutive scans. The hysteretic behaviour is reproducible for each single scan. The time from one scan and the next is approximately 10 minutes.

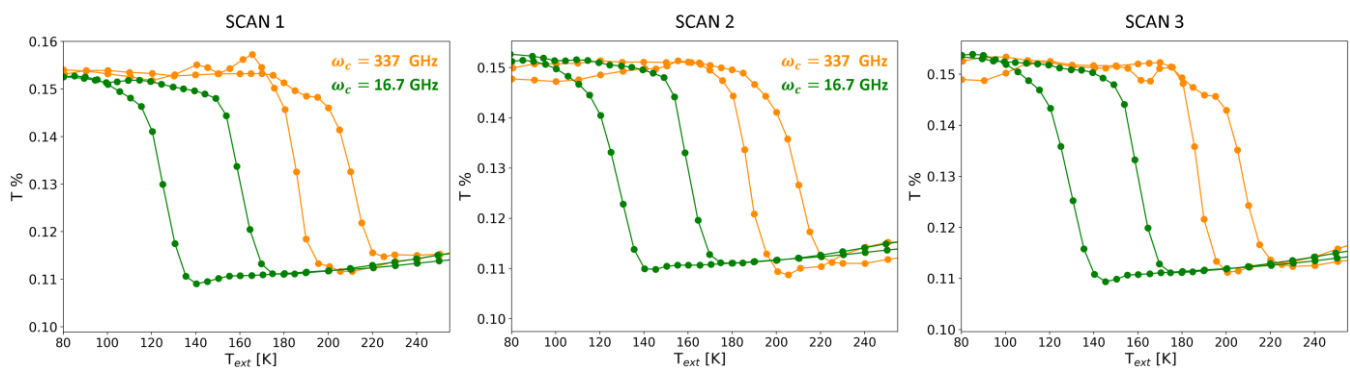

**Fig S21: Single scans as function of temperature.** Low frequency THz transmission as function of temperature (both heating and cooling) for two different cavity fundamental modes: 337 GHz (orange curves) and 16.7 GHz (green curves). The time from one scan and the next is approximately 10 minutes.

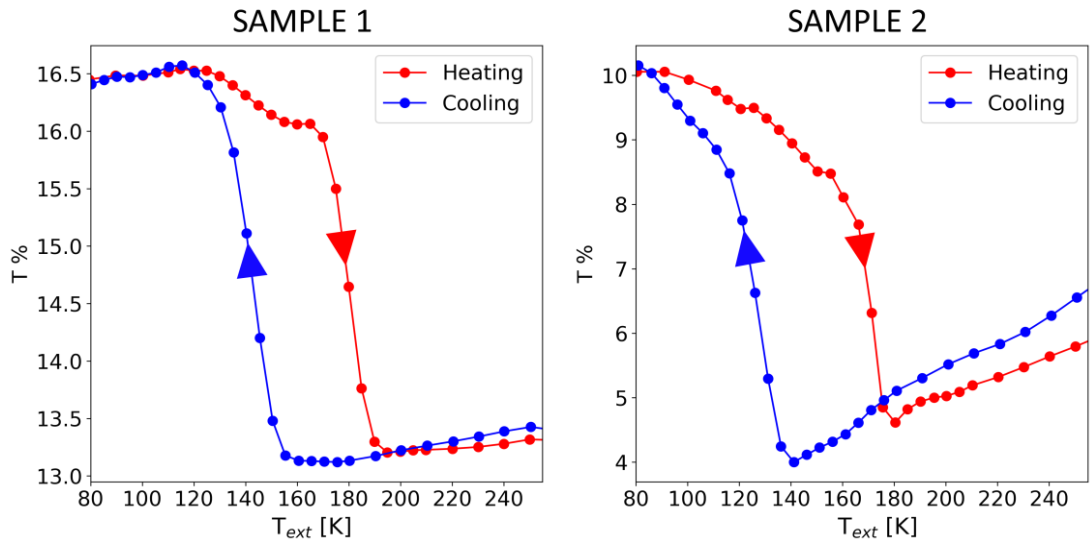

**Fig S22: Hysteretic behaviour in different samples.** Free-space hysteretic curves measured for two different samples belonging to the same batch. Sample 1 is the sample on which the measurements in the main manuscript have been performed.

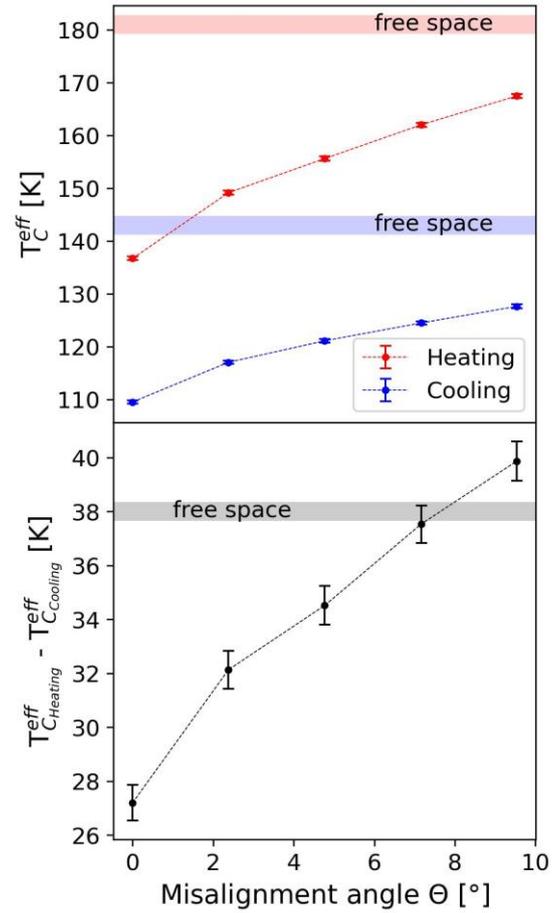

**Fig S23: Effective critical temperature as function of the cavity alignment.** Top panel: Effective critical temperature upon heating (red) and cooling (blue) the sample as function of the misalignment angle of the cavity. The shaded horizontal lines indicate the free-space reference. Bottom panel: corresponding effective hysteresis.